\documentclass[11pt]{article}

\usepackage{srcltx}

\usepackage{amsmath,amssymb,amsfonts,epsfig}
\usepackage{color}
\usepackage{subfigure}

\usepackage{fullpage}
\usepackage{setspace}
\usepackage{cite}
\usepackage{url}
\usepackage[normalem]{ulem}

\usepackage{graphicx}
\usepackage{graphics}
\usepackage{multirow}
\usepackage{multicol}
\usepackage{epstopdf}
\usepackage{epsfig} 
\usepackage{float}
\usepackage{booktabs}

\usepackage{flushend}

\newcommand{\tablesize}{\fontsize{9}{10}\selectfont}

\title{Improved 8-point Approximate DCT for Image and Video Compression Requiring
Only 14 Additions%
}

\author{%
Uma~Sadhvi~Potluri%
\thanks{%
Uma Sadhvi Potluri,
Arjuna Madanayake,
Sunera Kulasekera
and
Amila Edirisuriya
are with the
Department of Electrical and Computer Engineering,
The University of Akron, Akron, OH, USA
(e-mail: arjuna@uakron.edu).}
\quad
Arjuna~Madanayake$^\ast$
\\
Renato~J.~Cintra%
\thanks{Renato~J.~Cintra
is with the
Signal Processing Group,
Departamento de Estat\'istica,
Universidade Federal de Pernambuco, 
Recife, PE, Brazil
(e-mail: rjdsc@ieee.org).}
\quad
F\'abio~M.~Bayer%
\thanks{%
F\'abio~M.~Bayer
is with the
Departamento de Estat\'istica 
and Laborat\'orio de Ci\^encias Espaciais de Santa Maria (LACESM),
Universidade Federal de Santa Maria, 
Santa Maria, RS, Brazil
(e-mail: bayer@ufsm.br).}
\\
Sunera~Kulasekera$^\ast$
\quad
Amila~Edirisuriya$^\ast$
}

\date{}

\begin{document}

\onehalfspacing

\maketitle

\begin{abstract}

Video processing systems such as HEVC requiring low energy consumption needed for the multimedia market has lead to extensive development in fast algorithms for the efficient approximation
of \mbox{2-D} DCT transforms.
The DCT is  employed in a multitude of compression standards 
due to its remarkable energy compaction properties.
Multiplier-free approximate DCT transforms have been proposed that offer superior
compression performance at very low circuit complexity.
Such approximations
can be realized in digital VLSI hardware
using additions and subtractions only, leading to significant reductions in chip area and power consumption compared to conventional DCTs and integer transforms.
In this paper, 
we introduce a novel 8-point DCT approximation that requires \emph{only} 14 addition operations
and no multiplications.
The proposed transform possesses low computational complexity 
and 
is compared to state-of-the-art
DCT approximations in terms of both algorithm complexity and 
peak
signal-to-noise ratio.
The proposed DCT approximation is a candidate for reconfigurable video standards such as HEVC.
The proposed transform and several other DCT approximations are mapped 
to systolic-array digital architectures and physically realized as digital prototype circuits using
FPGA technology and mapped to 45~nm CMOS technology.

\end{abstract}

\begin{center}
\end{center}

\section{Introduction}

Recent years have experienced
a significant demand for
high dynamic range systems 
that 
operate at high resolutions~\cite{V5}.
In particular,
high-quality digital video in multimedia devices~\cite{multimedia} 
and
video-over-Internet protocol networks~\cite{video}
are prominent areas where
such requirements are evident. 
Other noticeable 
fields 
are
geospatial remote sensing~\cite{V3}, 
traffic cameras~\cite{V4}, 
automatic surveillance~\cite{V5}, 
homeland security~\cite{V6}, 
automotive industry~\cite{V7},
and multimedia wireless sensor networks~\cite{V1},
to name but a few.
Often
hardware capable of significant throughput is necessary;
as well as
allowable area-time complexity~\cite{V1}.

In this context,
the discrete cosine transform
(DCT)~\cite{ahmed1974, rao1990discrete, britanak2007discrete}
is an essential mathematical tool 
in both image and video 
coding~\cite{V1,bhaskaran1997, britanak2007discrete, Liang2001, haweel2001square, lengwehasatit2004scalable}. 
Indeed,
the DCT was demonstrated
to provide good energy compaction for natural images,
which can be described by first-order 
Markov signals~\cite{rao1990discrete, britanak2007discrete, Liang2001}.
Moreover,
in many situations,
the DCT is a very close substitute 
for the 
Karhunen-Lo\`eve transform (KLT),
which has optimal properties~\cite{ahmed1974, 
Clarke1981, 
rao1990discrete, 
britanak2007discrete, 
Liang2001, 
haweel2001square}.
As a result,
the two-dimensional (2-D) version of the
8-point DCT 
was adopted in several imaging standards
such as 
JPEG~\cite{penn1992}, 
MPEG-1~\cite{roma2007hybrid}, 
MPEG-2~\cite{mpeg2}, 
H.261~\cite{h261}, 
H.263~\cite{h263,V2},
and
H.264/AVC~\cite{h264,Wiegand2003}.

Additionally,
new compression schemes such as the High Efficiency Video Coding (HEVC) 
employs
DCT-like integer transforms
operating at
various block sizes ranging from 4$\times$4 to 32$\times$32 
pixels~\cite{hevc1,Park2012,Ohm2012}.
The distinctive characteristic of HEVC is
its capability of
achieving high compression performance
at
approximately half the bit rate required by
H.264/AVC with same image quality~\cite{hevc1,Park2012,Ohm2012}.
Also
HEVC 
was demonstrated to be 
especially effective for high-resolution video applications~\cite{Ohm2012}.
However, 
HEVC possesses a significant computational 
complexity
in terms of arithmetic operations~\cite{Park2012,Ohm2012,sullivan2012overview}.
In fact,
HEVC can be 2--4 times more computationally demanding
when compared to H.264/AVC~\cite{Park2012}.
Therefore, low complexity DCT-like approximations may benefit future video codecs including emerging
HEVC/H.265 systems.

Several efficient algorithms 
were developed and
a noticeable literature is available~\cite{rao1990discrete,wang1984fast, lee1984new, vetterli1984simple, hou1987fast, arai1988fast, loeffler1989practical, fw1992}.
Although 
fast algorithms
can significantly reduce
the computational complexity
of computing the DCT, 
floating-point operations are still required~\cite{britanak2007discrete}.
Despite their accuracy,
floating-point operations
are expensive in terms of circuitry complexity and power consumption.
Therefore,
minimizing
the number of floating-point operations
is a sought property
in a fast algorithm.
One way of circumventing this issue
is by means of approximate transforms.

The aim of this paper
is two-fold.
First,
we introduce %
a new DCT approximation
that possesses an extremely
low arithmetic complexity,
\emph{requiring only 14 additions}.
This novel transform
was obtained by means of
solving a tailored optimization problem
aiming at minimizing the
transform computational cost.
Second, 
we 
propose hardware implementations
for several 
\mbox{2-D} 8-point approximate DCT.
The 
approximate DCT methods under consideration
are
(i)~the proposed transform;
(ii)~the 2008 Bouguezel-Ahmad-Swamy (BAS) DCT approximation~\cite{bouguezel2008low};
(iii)~the parametric transform for image compression~\cite{bouguezel2011parametric};
(iv)~the Cintra-Bayer (CB) approximate DCT 
based on the rounding-off function~\cite{cintra2011dct};
(v)~the modified CB approximate DCT~\cite{cb14};
and
(vi)~the DCT approximation proposed in~\cite{multibeam2012}
in the context of beamforming.
All introduced implementations
are sought to be fully parallel time-multiplexed \mbox{2-D} architectures 
for 8$\times$8 data blocks.
Additionally,
the
proposed designs
are based on successive calls of 
\mbox{1-D} architectures 
taking advantage of
the separability property of the 
\mbox{2-D} DCT kernel.
Designs
were thoroughly
assessed
and compared.

This paper unfolds as follows.
In Section~\ref{sec:app}, we discuss the role of DCT-like fast algorithms for video CODECs while
proposing some new possibilities for low-power video processing where rapid reconfiguration of the hardware realization is possible.
In Section~\ref{section-review},
we review selected approximate methods for DCT computation
and describe associate fast algorithms in terms of matrix factorizations.
Section~\ref{section-proposed} 
details the proposed transform and its fast algorithm based on matrix factorizations.
Section~\ref{section-performance} 
discusses the computational complexity of the approximate DCT techniques.
Performance measures are also quantified and evaluated to assess
the proposed approximate DCT as well as the remaining selected approximations.
In Section~\ref{section-hardware} 
digital hardware architectures for discussed algorithms
are supplied both for \mbox{1-D} and \mbox{2-D} analysis.
Hardware resource consumptions using 
field programmable gate array (FPGA)
and CMOS 45~nm application-specific integrated circuit (ASIC)
technologies are tabulated.
Conclusions and final remarks
are in Section~\ref{section-conclusion}.

\section{Reconfigurable DCT-like Fast Algorithms in Video CODECs}
\label{sec:app}

In current literature,
several
approximate methods for the %
DCT calculation
have been archived~\cite{britanak2007discrete}.
While not computing the DCT
exactly, such approximations can provide 
meaningful estimations at low-complexity requirements.
In particular,
some DCT approximations
can totally eliminate
the requirement for floating-point operations---all calculations
are performed over a fixed-point arithmetic framework.
Prominent
8-point
approximation-based techniques were proposed in~\cite{haweel2001square,lengwehasatit2004scalable,
bouguezel2008low,bouguezel2008multiplication, bouguezel2009fast, bouguezel2010novel,bouguezel2011parametric,
bayer2010compression,cintra2011dct,cb14,multibeam2012}.
Works addressing 16-point DCT approximations
are also archived in 
literature~\cite{bouguezel2010novel,Bayer201216-point,Edirisuriya2012}.

In general, 
these approximation methods employ a transformation matrix 
whose elements are defined over the set $\{ 0, \pm 1/2, \pm1, \pm2\}$.
This implies null multiplicative complexity,
because the required operations
can be implemented 
exclusively by means of 
binary additions and shift operations.
Such DCT approximations 
can provide low-cost and low-power 
designs
and effectively 
replace the \emph{exact} DCT 
and other DCT-like transforms.
Indeed,
the performance characteristics of the low complexity DCT approximations 
appear similar to the exact DCT,
while their associate hardware implementations
are economical because of the absence of 
multipliers~\cite{haweel2001square,lengwehasatit2004scalable,
bouguezel2008low,bouguezel2008multiplication, bouguezel2009fast,
bouguezel2010novel,bouguezel2011parametric,
bayer2010compression,cintra2011dct,cb14,multibeam2012,bouguezel2010novel,
Bayer201216-point,Edirisuriya2012}. 
As a consequence,
some prospective applications of DCT approximations are 
found in
real-time video transmission and processing. 

Emerging video standards such as HEVC provide for reconfigurable operation on-the-fly which makes the availability of an ensemble of fast algorithms and digital VLSI architectures a valuable asset for low-energy high-performance embedded systems. 
For certain applications, low circuit complexity and/or power consumption is the driving factor, while for certain other applications, highest picture quality for reasonably low power consumption and/or complexity may be more important. 
In emerging systems, it may be possible to switch \emph{modus operandi} based on the demanded picture quality vs available energy in the device.
Such feature would be invaluable in high quality smart video devices demanding extended battery life.
Thus,
the availability of a suite of fast algorithms and implementation libraries for several efficient DCT approximation algorithms may be a welcoming contribution.

For example, in a future HEVC system, it may be possible to reconfigure the DCT engine to use a higher complexity DCT approximation which offers better signal-to-noise ratio (SNR) when the master device is powered by a remote power source, and then have the device seamlessly switch into a low complexity fast DCT algorithm when the battery storage falls below a certain threshold, for example~\cite{hevcdct}. 
Alternatively, the CODEC may be reconfigured in real-time to switch between different DCT approximations offering varying picture quality and power consumptions depending on the measured SNR of the incoming video stream, which would be content specific and very difficult to predict without resorting to real-time video metrics~\cite{hevcohm}.

Furthermore, another possible application for a suite of DCT approximation algorithms in the light of reconfigurable video codecs is the intelligent intra-frame fast reconfiguration of the DCT core to take into account certain local frame information and measured SNR metrics. 
For example, certain parts of a frame can demand better picture quality (foreground, say) when compared to relatively unimportant part of the frame (background, say)~\cite{hevcohm}. 
In such a case, it may be possible to switch DCT approximations algorithms on an intra frame basis to take into account the varying demands for picture clarity within a frame as well as the availability of reconfigurable logic based digital DCT engines that support fast reconfiguration in real-time.

\section{Review of Approximate DCT Methods}
\label{section-review}

In this section,
we review the mathematical description
of the selected 8-point DCT approximations. 
All 
discussed methods here
consist of a transformation matrix 
that can be put in the following
format:
\begin{align*}
[\text{diagonal matrix}]
\times
[\text{low-complexity matrix}]
.
\end{align*}
The diagonal matrix usually contains
irrational numbers in the form $1/\sqrt{m}$,
where $m$ is a small positive integer.
In principle,
the irrational numbers required 
in the diagonal matrix
would require an increased computational complexity.
However,
in the context of image compression,
the diagonal matrix can simply be
absorbed into the quantization step
of JPEG-like compression procedures~\cite{
lengwehasatit2004scalable,
bouguezel2008low,
bouguezel2009fast,
bouguezel2011parametric,
cintra2011dct, 
cb14}. 
Therefore,
in this case,
the complexity of the approximation
is bounded by the complexity of
the low-complexity matrix.
Since the entries of the low complexity matrix
comprise only
powers of two
in 
$\{ 0, \pm 1/2, \pm1, \pm2\}$,
null multiplicative complexity.
is achieved.

In the next subsections,
we detail these methods in terms of its
transformation matrices
and the associated fast algorithms
obtained by matrix factorization techniques.
All derived
fast algorithms employ sparse matrices
whose elements are the above-mentioned
powers of two.

\subsection{Bouguezel-Ahmad-Swamy Approximate DCT}

In~\cite{bouguezel2008low},
a low-complexity approximate was introduced
by Bouguezel~\emph{et al.}
We refer to this approximate DCT
as BAS-2008 approximation.
The BAS-2008 approximation~$\mathbf{C}_1$
has
the following mathematical structure:
\begin{align*}
\mathbf{C_1}
=
\mathbf{D_1}
\cdot
\mathbf{T_1}
=
\mathbf{D_1}
\cdot
\begin{bmatrix}
\begin{smallmatrix}
1   &  1   &  1   &  1   &  1   &  1   &  1   & 1 \\
1   &  1   &  0   &  0   &  0   &  0   & -1   & -1 \\
1   &  \frac{1}{2} & -\frac{1}{2} & -1   & -1   & -\frac{1}{2} &  \frac{1}{2} & 1 \\
0   &  0   & -1   &  0   &  0   &  1   &  0   & 0 \\
1   & -1   & -1   &  1   &  1   & -1   & -1   & 1 \\
1   & -1   &  0   &  0   &  0   &  0   &  1   & -1 \\
\frac{1}{2} & -1   &  1   & -\frac{1}{2} & -\frac{1}{2} &  1   & -1   & \frac{1}{2} \\
0   &  0   &  0   & -1   &  1   &  0   &  0   & 0 
\end{smallmatrix}
\end{bmatrix},
\end{align*}

\noindent
where $\mathbf{D}_1 = 
\operatorname{diag}\left( 
\frac{1}{\sqrt{8}},
\frac{1}{\sqrt{4}},
\frac{1}{\sqrt{5}},
\frac{1}{\sqrt{2}},
\frac{1}{\sqrt{8}},
\frac{1}{\sqrt{4}},
\frac{1}{\sqrt{5}},
\frac{1}{\sqrt{2}}
\right)$. 
A fast algorithm for 
matrix
$\mathbf{T_1}$ can be derived by means of matrix factorization.
Indeed,
$\mathbf{T_1}$
can be written as 
a product of three sparse matrices having 
$\{0, \pm 1/2, \pm 1\}$ elements
as shown below~\cite{bouguezel2008low}:
$
\mathbf{T_1} 
=
\mathbf{A_3} \cdot \mathbf{A_2} \cdot \mathbf{A_1}
$,
where 
$
\mathbf{A_1}
=
\begin{bmatrix}
\begin{smallmatrix}
\mathbf{I}_4 & \bar{\mathbf{I}}_4 \\
\bar{\mathbf{I}}_4 & - \mathbf{I}_4 \\
\end{smallmatrix}
\end{bmatrix}
$,
\begin{align*}
\mathbf{A_2}
=
\begin{bmatrix}
\begin{smallmatrix}
1 & 0 & 0 & 1 & 0 &0 & 0 & 0 \\
0  &  0   &  0   &  0   &  0   &  0   & 1   & 1 \\
0   & 1 & 1 & 0   & 0   & 0 &  0 & 0 \\
0   &  0   & 0  &  0   &  0  &  -1  &  0   & 0 \\
0   & 1   & -1  &  0   &  0  & 0   & 0   & 0 \\
0   & 0   &  0   &  0   &  0   &  0   &  -1   & 1 \\
1 & 0   &  0   & -1 & 0 & 0   & 0   & 0 \\
0   &  0   &  0   & 0  & -1   &  0   &  0   & 0\\ 
\end{smallmatrix}
\end{bmatrix}
,
\mathbf{A_3}
&=
\begin{bmatrix}
\begin{smallmatrix}
1 & 0 & 1 & 0 & 0 & 0 &  0   & 0 \\
0  &  1   &  0   &  0   &  0   &  0   & 0   & 0 \\
0  & 0 & 0 & 0  & \frac{1}{2} & 0 & 1 & 0 \\
0   &  0   & 0  &  1   &  0  &  0  &  0   & 0 \\
1 & 0   & -1  &  0   &  0  & 0   & 0   & 0 \\
0   & 0   &  0   &  0   &  0   &  1   &  0   & 0 \\
0 & 0   &  0   & 0 & -1 & 0   & \frac{1}{2}   & 0 \\
0   &  0   &  0   & 0  & 0   &  0   &  0   & 1\\ 
\end{smallmatrix}
\end{bmatrix}
.
\end{align*}
Matrices $\mathbf{I}_n$ and $\bar{\mathbf{I}}_n$
denote the identity and counter-identity
matrices of order $n$,
respectively.
It is recognizable that
matrix~$\mathbf{A_1}$ 
is the well-known decimation-in-frequency structure 
present in several fast algorithms~\cite{britanak2007discrete}.

\subsection{Parametric Transform}

Proposed in 2011
by Bouguezel-Ahmad-Swamy~\cite{bouguezel2011parametric},
the parametric transform
is
an 8-point orthogonal transform
containing
a single parameter~$a$
in its transformation matrix $\mathbf{C}^{(a)}$.
In this work,
we refer to this method as 
the \mbox{BAS-2011} transform.
It is given as follows:
\begin{align*}
\mathbf{C}^{(a)}
=
\mathbf{D}^{(a)}
\cdot
\mathbf{T}^{(a)}
=
\mathbf{D}^{(a)}
\cdot
\begin{bmatrix}
\begin{smallmatrix}
1   &  1   &  1   &  1   &  1   &  1   &  1   & 1 \\
1   &  1   &  0   &  0   &  0   &  0   & -1   & -1 \\
1   &  a   &  -a   &  -1   &  -1   &  -a   &  a   & 1 \\
0   &  0   &  1   &  0   &  0   &  -1   &  0   & 0 \\
1   &  -1   &  -1   &  1   &  1   &  -1   &  -1   & 1 \\
0   &  0   &  0   &  1   &  -1   &  0   &  0   & 0 \\
1   &  -1   &  0   &  0   &  0   &  0   &  1   & -1 \\
a   &  -1   &  1   &  -a   &  -a   &  1   &  -1   & a
\end{smallmatrix}
\end{bmatrix}
, 
\end{align*}
\noindent
where $\mathbf{D}^{(a)} \!\!= \!\!
\operatorname{diag}\left( 
\frac{1}{\sqrt{8}},
\frac{1}{2},
\frac{1}{\sqrt{4+4a^2}},
\frac{1}{\sqrt{2}},
\frac{1}{\sqrt{8}},
\frac{1}{\sqrt{2}},
\frac{1}{2},
\frac{1}{\sqrt{4+4a^2}}
\right)$.
Usually
the parameter~$a$ is selected as a small integer
in order to minimize the complexity of~$\mathbf{T}^{(a)}$.
In~\cite{bouguezel2011parametric},
suggested values are $a\in\{0,1/2,1\}$.
The value $a=1/2$ will not be considered in our analyses
because
in hardware it represents a right-shift
which may incur in computational errors.
Another possible value that furnishes 
a low-complexity, error-free transform
is $a = 2$.
The matrix factorization
of $\mathbf{T}^{(a)}$ 
that leads to its fast algorithm
is~\cite{bouguezel2011parametric}:
$
\mathbf{T}^{(a)}
=
\mathbf{P_1}
\cdot
\mathbf{Q}^{(a)}
\cdot
\mathbf{A_4}
\cdot
\mathbf{A_1}
$,
\noindent
where
$
\mathbf{Q}^{(a)}
=
\operatorname{diag}
\left(
\begin{bmatrix}
\begin{smallmatrix}
1 & 1\\
1 & -1
\end{smallmatrix}
\end{bmatrix}
,
\begin{bmatrix}
\begin{smallmatrix}
a & 1\\
-1 & a
\end{smallmatrix}
\end{bmatrix}
,
\mathbf{I}_4
\right)
,
$
and
$
\mathbf{A_4}=
\operatorname{diag}
\left(
\begin{bmatrix}
\begin{smallmatrix}
1 &0 &0 &1 \\
0 &1 &1 &0 \\
0 &1 &-1 &0\\
1 &0 &0 &-1\\
\end{smallmatrix}
\end{bmatrix}
,
\mathbf{I}_2,
\begin{bmatrix}
\begin{smallmatrix}
1 &1\\
-1 &1
\end{smallmatrix}
\end{bmatrix}
\right)
$.
Matrix~$\mathbf{P_1}$
performs the simple permutation 
(1)(2 5 6 4 8 7)(3),
where
cyclic notation is employed~\cite[p.~77]{Herstein1975}.
This is a compact notation to denote permutation.
In this particular case, 
it means that component indices
are permuted according to
$2\to5\to6\to4\to8\to7\to2$.
Indices 1 and 3 are unchanged.
Therefore,
$\mathbf{P_1}$ represents no computational complexity.

\subsection{\mbox{CB-2011} Approximation}

By means
of judiciously rounding-off
the elements of the exact DCT matrix,
a DCT approximation was obtained
and
described in~\cite{cintra2011dct}.
The resulting 8-point approximation matrix
is orthogonal
and contains only elements 
in $\{0, \pm 1\}$.
Clearly,
it possesses very low arithmetic complexity~\cite{cintra2011dct}.
The matrix derived transformation matrix~$\mathbf{C_2}$ 
is given by:
\begin{align*}
\mathbf{C_2}
=
\mathbf{D_2}
\cdot
\mathbf{T_2}
=
\mathbf{D_2}
\cdot
\begin{bmatrix}
\begin{smallmatrix}
1   &  1   &  1   &  1  &  1   &  1  &  1  &  1 \\
1   &  1   &  1   &  0  &  0   & -1  & -1  & -1 \\
1   &  0   &  0   & -1  & -1   &  0  &  0  &  1 \\
1   &  0   & -1   & -1  &  1   &  1  &  0  & -1 \\
1   & -1   & -1   &  1  &  1   & -1  & -1  &  1 \\
1   & -1   &  0   &  1  & -1   &  0  &  1  & -1 \\
0   & -1   &  1   &  0  &  0   &  1  & -1  &  0 \\
0   & -1   &  1   & -1  &  1   & -1  &  1  &  0
\end{smallmatrix}
\end{bmatrix}
,
\end{align*}
where $\mathbf{D_2} = 
\operatorname{diag}\left( 
\frac{1}{\sqrt{8}},
\frac{1}{\sqrt{6}},
\frac{1}{2},
\frac{1}{\sqrt{6}},
\frac{1}{\sqrt{8}},
\frac{1}{\sqrt{6}},
\frac{1}{2},
\frac{1}{\sqrt{6}}
\right)$.
An efficient factorization for
the fast algorithm for
$\mathbf{T_2}$
was proposed in~\cite{cintra2011dct}
as described below:
$
\mathbf{T_2}
= 
\mathbf{P_2}
\cdot
\mathbf{A_6} 
\cdot
\mathbf{A_5} 
\cdot 
\mathbf{A_1}
$,
where
$
\mathbf{A_5}
=
\operatorname{diag}
\left(
\begin{bmatrix}
\begin{smallmatrix}
1 & 0 &  0 &  1 \\
0 & 1 &  1 &  0 \\
0 & 1 & -1 &  0 \\
1 & 0 &  0 & -1 \\
\end{smallmatrix}
\end{bmatrix}
,
\begin{bmatrix}
\begin{smallmatrix}
-1 &  1 & -1 &  0 \\
-1 & -1 &  0 &  1 \\
 1 &  0 & -1 &  1 \\
 0 &  1 &  1 &  1 
\end{smallmatrix}
\end{bmatrix}
\right)
$
and
$
\mathbf{A_6}
=
\operatorname{diag}
\left(
\begin{bmatrix}
\begin{smallmatrix}
1 &  1 \\
1 & -1 \\
\end{smallmatrix}
\end{bmatrix}
,
-1,
\mathbf{I}_5
\right)
$.
Matrix~$\mathbf{P_2}$ corresponds to the following
permutation: (1)(2 5 8)(3 7 6 4).

\subsection{Modified \mbox{CB-2011} Approximation}

The transform proposed in~\cite{cb14}
is obtained by replacing elements of the \mbox{CB-2011} matrix with zeros.
The resulting matrix
is given by:
\begin{align*}
\mathbf{C_3}
=
\mathbf{D_3}
\cdot
\mathbf{T_3} 
=
\mathbf{D_3}
\cdot
\begin{bmatrix}
\begin{smallmatrix}
1 & 1 & 1 &  1 &  1 & 1 & 1 & 1 \\
1 & 0 & 0 &  0 &  0 & 0 & 0 &-1 \\
1 & 0 & 0 &-1 & -1 & 0 & 0 & 1 \\
0 & 0 &-1 & 0  &  0 & 1 & 0 & 0 \\
1 &-1 &-1 &  1 &  1 &-1 &-1 & 1 \\
0 &-1 & 0 &  0 &  0 & 0 & 1 & 0 \\
0 &-1 & 1 &  0 &  0 & 1 &-1 & 0 \\
0 & 0 & 0 &-1 &  1 & 0 & 0 & 0
\end{smallmatrix}
\end{bmatrix}
, 
\end{align*}  
where 
$
\mathbf{D_3} 
= 
\operatorname{diag}
\left(
\frac{1}{\sqrt{8}},
\frac{1}{\sqrt{2}},
\frac{1}{2},
\frac{1}{\sqrt{2}},
\frac{1}{\sqrt{8}},
\frac{1}{\sqrt{2}},
\frac{1}{2},
\frac{1}{\sqrt{2}}
\right)
$. 
Matrix $\mathbf{T_3}$
can be factorized into
$
\mathbf{T_3}= 
\mathbf{P_2} \cdot 
\mathbf{A_6} \cdot 
\mathbf{A_7} \cdot 
\mathbf{A_1}
$,
where 
$
\mathbf{A_7} =
\operatorname{diag}
\left(
\begin{bmatrix}
\begin{smallmatrix}
1 & 0 & 0 &  1 \\
0 & 1 & 1 &  0 \\
0 & 1 & -1&  0 \\
1 & 0 & 0 & -1 &
\end{smallmatrix}
\end{bmatrix},
-\mathbf{I}_3,
1
\right)
$.
This particular DCT approximation
has the distinction of requiring only 14~additions
for its computation~\cite{cb14}.

\subsection{Approximate DCT in~\cite{multibeam2012}}

In~\cite{multibeam2012},
a DCT approximation tailored
for a particular radio-frequency (RF) application
was obtained in accordance with
an exhaustive computational search.
This transformation is given by
\begin{align*}
\mathbf{C_4}
=
\mathbf{D_4}
\cdot
\mathbf{T_4}
=
\mathbf{D_4}
\cdot
\begin{bmatrix}
\begin{smallmatrix}
1  &  1 &  1  &  1  &  1  &  1  &  1  &  1 \\
2  &  1 &  1  &  0  &  0  & -1  & -1  & -2 \\
2  &  1 & -1  & -2  & -2  & -1  &  1  &  2 \\
1  &  0 & -2  & -1  &  1  &  2  &  0  & -1 \\
1  & -1 & -1  &  1  &  1  & -1  & -1  &  1 \\
1  & -2 &  0  &  1  & -1  &  0  &  2  & -1 \\
1  & -2 &  2  & -1  & -1  &  2  & -2  &  1 \\
0  & -1 &  1  & -2  &  2  & -1  &  1  &  0 \\
\end{smallmatrix}
\end{bmatrix}
,
\end{align*}
where $\mathbf{D_4} = 
\frac{1}{2}
\cdot
\operatorname{diag}
\left( 
\frac{1}{\sqrt{2}},
\frac{1}{\sqrt{3}},
\frac{1}{\sqrt{5}},
\frac{1}{\sqrt{3}},
\frac{1}{\sqrt{2}},
\frac{1}{\sqrt{3}},
\frac{1}{\sqrt{5}},
\frac{1}{\sqrt{3}}
\right)$. 
The fast algorithm for its computation
consists of the following
matrix factorization:
$
\mathbf{T_4}
=
\mathbf{P_3} \cdot
\mathbf{A_9} \cdot
\mathbf{A_8} \cdot
\mathbf{A_1}
$,
where
$
\mathbf{A_9}
=
\operatorname{diag}
\left(
\begin{bmatrix}
\begin{smallmatrix}
    1  &  1 \\
    1  & -1
\end{smallmatrix}
\end{bmatrix}
,
\begin{bmatrix}
\begin{smallmatrix}
 1 &   2 \\
-2 &   1 
\end{smallmatrix}
\end{bmatrix}
,
\mathbf{I}_4
\right)
$,
$
\mathbf{A_8}
=
\operatorname{diag}
\left(
\begin{bmatrix}
\begin{smallmatrix}
    1  &  0 &   0 &   1 \\
    0  &  1 &   1 &   0 \\
    0  &  1 &  -1 &   0 \\
    1  &  0 &   0 &  -1 \\
\end{smallmatrix}
\end{bmatrix}
,
\begin{bmatrix}
\begin{smallmatrix}
 0 &   1 &   1 &   2 \\
-1 &  -2 &   0 &   1 \\
 1 &   0 &  -2 &   1 \\
-2 &   1 &  -1 &   0 
\end{smallmatrix}
\end{bmatrix}
\right)
$,
and
matrix~$\mathbf{P_3}$
denotes the permutation (1)(2 5)(3)(4 7 6)(8).

\section{Proposed Transform}
\label{section-proposed}

We aim at deriving a novel low-complexity approximate DCT.
For such end,
we propose
a search over the $8\times 8$
matrix space
in order to find candidate matrices
that possess low computation cost.
Let us define the cost of a transformation
matrix
as the number of arithmetic operations
required for its computation.
One way to guarantee
good candidates is to restrict the search to
matrices whose entries
do not require multiplication operations.
Thus we have the following optimization problem:
\begin{align}
\label{equation-optimization}
\mathbf{T}^ \ast
=
\arg
\min_\mathbf{T}
\operatorname{cost}(\mathbf{T}),
\end{align}
where $\mathbf{T}^\ast$ is the sought matrix
and
$\operatorname{cost}(\mathbf{T})$ 
returns the arithmetic complexity of~$\mathbf{T}$.
Additionally,
the following constraints were adopted:
\begin{enumerate}
\item  Elements of matrix~$\mathbf{T}$
must be in $\{0, \pm1, \pm2\}$ 
to ensure that resulting multiplicative complexity is null;

\item
We impose the following form for matrix~$\mathbf{T}$:
\begin{align*}
\!\!\!\!\!\!
\mathbf{T} 
\!=\!\!
\begin{bmatrix}
\begin{smallmatrix}
a_3 & \phantom{-}a_3 & \phantom{-}a_3 & \phantom{-}a_3 & \phantom{-}a_3 & \phantom{-}a_3 & \phantom{-}a_3 & \phantom{-}a_3 \\
a_0 & \phantom{-}a_2 & \phantom{-}a_4 & \phantom{-}a_6 & -a_6 & -a_4 & -a_2 & -a_0 \\
a_1 & \phantom{-}a_5 & -a_5 & -a_1 & -a_1 & -a_5 & \phantom{-}a_5 & \phantom{-}a_1 \\
a_2 & -a_6 & -a_0 & -a_4 & \phantom{-}a_4 & \phantom{-}a_0 & \phantom{-}a_6 & -a_2 \\
a_3 & -a_3 & -a_3 & \phantom{-}a_3 & \phantom{-}a_3 & -a_3 & -a_3 & \phantom{-}a_3 \\
a_4 & -a_0 & \phantom{-}a_6 & \phantom{-}a_2 & -a_2 & -a_6 & \phantom{-}a_0 & -a_4 \\
a_5 & -a_1 & \phantom{-}a_1 & -a_5 & -a_5 & \phantom{-}a_1 & -a_1 & \phantom{-}a_5 \\
a_6 & -a_4 & \phantom{-}a_2 & -a_0 & \phantom{-}a_0 & -a_2 & \phantom{-}a_4 & -a_6
\end{smallmatrix}
\end{bmatrix}
,
\end{align*}
where $a_i\in\{0, 1, 2\}$, 
for $i=0,1\ldots,6$;

\item 
All rows of $\mathbf{T}$ are non-null;

\item
Matrix $\mathbf{T}\cdot\mathbf{T}^{\top}$
must be a diagonal matrix
to ensure orthogonality of 
the resulting approximation~\cite{cintra2011integer}.

\end{enumerate}

Constraint~2)
is required 
to preserve
the DCT-like matrix structure.
We recall
that
the \emph{exact} 8-point DCT matrix 
is given by~\cite{fw1992}:
\begin{align*}
\label{dct8}
\mathbf{C}  = 
\frac{1}{2}
\cdot
\begin{bmatrix}
\begin{smallmatrix}
\gamma_3 & \phantom{-}\gamma_3 & \phantom{-}\gamma_3 & \phantom{-}\gamma_3 & \phantom{-}\gamma_3 & \phantom{-}\gamma_3 & \phantom{-}\gamma_3 & \phantom{-}\gamma_3 \\
\gamma_0 & \phantom{-}\gamma_2 & \phantom{-}\gamma_4 & \phantom{-}\gamma_6 & -\gamma_6 & -\gamma_4 & -\gamma_2 & -\gamma_0 \\
\gamma_1 & \phantom{-}\gamma_5 & -\gamma_5 & -\gamma_1 & -\gamma_1 & -\gamma_5 & \phantom{-}\gamma_5 & \phantom{-}\gamma_1 \\
\gamma_2 & -\gamma_6 & -\gamma_0 & -\gamma_4 & \phantom{-}\gamma_4 & \phantom{-}\gamma_0 & \phantom{-}\gamma_6 & -\gamma_2 \\
\gamma_3 & -\gamma_3 & -\gamma_3 & \phantom{-}\gamma_3 & \phantom{-}\gamma_3 & -\gamma_3 & -\gamma_3 & \phantom{-}\gamma_3 \\
\gamma_4 & -\gamma_0 & \phantom{-}\gamma_6 & \phantom{-}\gamma_2 & -\gamma_2 & -\gamma_6 & \phantom{-}\gamma_0 & -\gamma_4 \\
\gamma_5 & -\gamma_1 & \phantom{-}\gamma_1 & -\gamma_5 & -\gamma_5 & \phantom{-}\gamma_1 & -\gamma_1 & \phantom{-}\gamma_5 \\
\gamma_6 & -\gamma_4 & \phantom{-}\gamma_2 & -\gamma_0 & \phantom{-}\gamma_0 & -\gamma_2 & \phantom{-}\gamma_4 & -\gamma_6
\end{smallmatrix}
\end{bmatrix},
\end{align*}
where
$\gamma_k = \cos(2\pi (k+1) /32)$,
$k=0,1,\ldots,6$.

Above optimization problem is algebraically intractable.
Therefore we resorted to exhaustive computational search.
As a result,
eight candidate matrices were found, 
including the transform matrix proposed in~\cite{cb14}. 
Among these minimal cost matrices,
we separated the matrix 
that presents the best performance in 
terms of image quality of compressed images
according the JPEG-like technique 
employed in~\cite{bouguezel2008low, bouguezel2008multiplication, bouguezel2009fast, bouguezel2010novel, bouguezel2011parametric,
bayer2010compression, cintra2011dct, cb14}, 
and briefly 
reviewed
in next Section~\ref{section-performance}.

An important parameter in the image compression routine
is the number of retained coefficients
in the transform domain.
In several applications,
the number of retained coefficients is very low.
For instance, 
considering 8$\times$8 image blocks,
(i)~in image compression using support vector machine,
only the first 8--16 coefficients
were considered~\cite{Robinson2003};
(ii)~Mandyam~\emph{et~al.} 
proposed a method for image reconstruction
based on only three coefficients;
and
Bouguezel~\emph{et~al.}
employed
only 10 DCT coefficients
when assessing image compression
methods~\cite{bouguezel2008multiplication, bouguezel2009fast}.
Retaining a very small number of coefficients
is also common for other image block sizes.
In high speed face recognition applications,
Pan~\emph{et~al.}
demonstrated that just 0.34\%--24.26\% out of 92$\times$112 DCT
coefficients are sufficient~\cite{Pan1999,Pan2001}.
Therefore,
as a compromise,
we adopted the
number of retained coefficients
equal to~10,
as suggested in the experiments
by~Bouguezel~\emph{et~al.}~\cite{bouguezel2008multiplication, bouguezel2009fast}.

The solution of~\eqref{equation-optimization}
is the following
DCT approximation:
\begin{align*}
\mathbf{C}^\ast
=
\mathbf{D}^\ast
\cdot
\mathbf{T}^\ast
=
\mathbf{D}^\ast
\cdot
\begin{bmatrix}
\begin{smallmatrix}
1 &1 &1 & 1 & 1 &1 &1 &1 \\
0 &1 &0 & 0 & 0 &0 & -1 &0 \\
1 &0 &0 &-1 & -1&0 &0 &1 \\
1 &0 &0 & 0 & 0 &0 &0 &-1 \\
1 &-1&-1& 1 & 1 &-1&-1&1 \\
0 &0 &0 & 1 & -1 &0 &0 &0 \\
0 &-1&1 & 0 & 0 &1 &-1 &0 \\
0 &0 &1 & 0 & 0 &- 1 &0 &0
\end{smallmatrix}
\end{bmatrix}
,
\end{align*}  
where 
$
\mathbf{D}^\ast
=
\operatorname{diag}
\left(
\frac{1}{\sqrt{8}},
\frac{1}{\sqrt{2}},
\frac{1}{2},
\frac{1}{\sqrt{2}},
\frac{1}{\sqrt{8}},
\frac{1}{\sqrt{2}},
\frac{1}{2},
\frac{1}{\sqrt{2}}
\right)
$. 
Matrix 
$\mathbf{T}^\ast$ 
has entries in $\{0, \pm1\}$
and
it can be given a sparse factorization
according to:
$
\mathbf{T}^\ast
= 
\mathbf{P_4}
\cdot 
\mathbf{A_{12}}
\cdot 
\mathbf{A_{11}}
\cdot 
\mathbf{A_1}
$,
where
$
\mathbf{A_{11}}
=
\operatorname{diag}
\left(
\begin{bmatrix}
\begin{smallmatrix}
    1  &  0 &   0 &   1 \\
    0  &  1 &   1 &   0 \\
    0  &  1 &  -1 &   0 \\
    1  &  0 &   0 &  -1
\end{smallmatrix}
\end{bmatrix}
,
\mathbf{I}_4
\right)
$,
$
\mathbf{A_{12}}
=
\operatorname{diag}
\left(
\begin{bmatrix}
\begin{smallmatrix}
    1  &  1 \\
    1  & -1 
\end{smallmatrix}
\end{bmatrix}
,
-1,
\mathbf{I}_5
\right)
$,
and
$\mathbf{P}_4$
is the permutation
(1)(2 5 6 8 4 3 7).

\section{Computational Complexity and Performance Analysis}
\label{section-performance}

The performance of the DCT approximations
is often
a trade-off between
accuracy 
and 
computational complexity of a given algorithm~\cite{britanak2007discrete}.
In this section,
we assess the computational complexity of the discussed methods
and objectively compare them.
Additionally,
we separate several performance measures
to quantify 
how ``close''
each approximation are;
and
to evaluate their performance 
as image compression tool.

\subsection{Arithmetic Complexity}

We adopt the arithmetic complexity
as figure of merit
for estimating the computational complexity.
The arithmetic complexity 
consists of the number of elementary arithmetic
operations 
(additions/subtractions, multiplications/divisions, and
bit-shift operations)
required to compute a given transformation.
In other words,
in all cases,
we focus our attention
to the low-complexity matrices:
$\mathbf{T_1}$,
$\mathbf{T}^{(a)}$,
$\mathbf{T_2}$,
$\mathbf{T_3}$,
$\mathbf{T_4}$,
and
the proposed
matrix
$\mathbf{T}^\ast$.
For instance,
in the context of image and video compression,
the complexity of the diagonal matrix
can be absorbed into the quantization step~\cite{lengwehasatit2004scalable,
bouguezel2008low,
bouguezel2009fast,
bouguezel2011parametric,
cintra2011dct, 
cb14};
therefore the diagonal matrix does not contribute
towards an increase of 
the arithmetic complexity~\cite{cintra2011dct,cb14}.

Because
all considered DCT approximations
have null multiplicative complexity,
we resort to comparing them
in terms of their arithmetic complexity
assessed by
the number of
additions/subtractions and bit-shift operations.
Table~\ref{table-complexity}
displays the obtained complexities.
We also include the complexity of
the \emph{exact}
DCT calculated
(i)~directly from definition~\cite{rao1990discrete}
and
(ii)~according to Arai fast algorithm for the exact DCT~\cite{arai1988fast}.

We derived 
a fast algorithm 
for the proposed transform,
employing \emph{only 14 additions}.
This is the same
very low-complexity 
exhibited by 
the Modified \mbox{CB-2011} approximation~\cite{cb14}.
To the best of our knowledge
these are  DCT approximations
offering the lowest arithmetic complexity
in literature.

\begin{table}%

\begin{center}
\caption{Arithmetic complexity analysis}
\label{table-complexity}
\begin{tabular}{lcccc} %
\toprule
Method & Mult & Add & Shifts & Total\\
\midrule  
Exact DCT (Definition)~\cite{rao1990discrete} & 64 & 56 &  0 & 120 \\
Arai algorithm (exact)~\cite{arai1988fast} & 5 & 29 &  0 & 34\\
BAS-2008~\cite{bouguezel2008low} & 0 & 18 & 2 & 20 \\
\mbox{BAS-2011}~\cite{bouguezel2011parametric} with $a=0$ & 0 & 16 & 0 & 16 \\
\mbox{BAS-2011}~\cite{bouguezel2011parametric} with $a=1$ & 0 & 18 & 0 & 18\\
\mbox{BAS-2011}~\cite{bouguezel2011parametric} with $a=2$ & 0 & 18 & 2 & 20\\
\mbox{CB-2011}~\cite{cintra2011dct} & 0 & 22 & 0 & 22\\
Modified \mbox{CB-2011}~\cite{cb14} & 0 & 14 & 0 & 14 \\
Approximate DCT in~\cite{multibeam2012} & 0 & 24 & 6 & 30 \\
Proposed transform & 0 & 14 & 0 & 14\\
\bottomrule
\end{tabular}
\end{center}
\end{table}

\subsection{Comparative Performance}

We employed three classes of 
assessment tools:
(i)~matrix proximity metrics
with respect to the exact DCT matrix;
(ii)~transform-related measures;
and
(iii)~image quality measures in
image compression.
For the first class of measures,
we adopted
the total error energy~\cite{cintra2011dct}
and
the mean-square error~(MSE)~\cite{Liang2001,britanak2007discrete}.
For transform performance evaluation,
we selected
the transform coding gain~($C_g$)~\cite{Liang2001,britanak2007discrete}
and
the transform efficiency~($\eta$)~\cite{Cham1989,britanak2007discrete}.
Finally,
for image quality assessment
we employed 
the peak SNR~(PSNR)~\cite{Salomon2007,Quan2012}
and
the universal quality index~(UQI)~\cite{wang2002universal}.
Next subsections
furnish a brief description of each of these measures.

\subsubsection{Matrix Proximity Metrics}

Let
$\hat{\mathbf{C}}$ be an approximate DCT matrix
and
$\mathbf{C}$ be the exact DCT matrix.
The total error energy is
an energy-based measure for
quantifying the 
``distance''
between
$\mathbf{C}$
and
$\hat{\mathbf{C}}$.
It is described as follows~\cite{cintra2011dct}.

Let $H_m(\omega; \mathbf{T})$ is the transfer function
of the $m$-th row of a given matrix~$\mathbf{T}$
as shown below:
\begin{align*}
H_m(\omega;\mathbf{T})
= \!\!
\sum_{n=1}^8 
t_{m,n}
\exp
\big\{\!
- \! \jmath \,(n-\!1) \, \omega
\big\}
,
\;\,
m=1,2,\ldots,8
,
\end{align*}
where 
$\jmath=\sqrt{-1}$ 
and $t_{m,n}$ is the $(m,n)$-th element of $\mathbf{T}$.
Then the row-wise error energy
related to the difference between
$\mathbf{C}$
and
$\hat{\mathbf{C}}$
is furnished by:
\begin{align*}
D_m(\omega;\hat{\mathbf{C}})
\triangleq
\left|
H_m(\omega;\mathbf{C}-\hat{\mathbf{C}})
\right|^2
,
\;\;
m=1,2,\ldots,8
.
\end{align*}
We note that,
for each row $m$ at any angular frequency~$\omega \in [0,\pi]$ 
in radians per sample,
$D_m(\omega; \mathbf{C}-\hat{\mathbf{C}} )$ expression
quantifies how discrepant
a given 
approximation 
matrix~$\hat{\mathbf{C}}$ is 
from the
matrix $\mathbf{C}$.
In this way, 
a total error energy
departing from the \emph{exact} DCT 
can be obtained by 
\cite{cintra2011dct}:
\begin{equation*}
\label{equation-total-error-energy}
\epsilon
=
\sum_{m=1}^8
\int_0^\pi
D_m(\omega;\hat{\mathbf{C}})
\,
\mathrm{d}\omega
.
\end{equation*}
Above integral can
be computed by 
numerical quadrature methods~\cite{piessens1983quadpack}.

For the MSE evaluation,
we assume that the input signal 
is a first-order Gaussian Markov process
with zero-mean, unit variance, and correlation equal to 0{.}95~\cite{Liang2001,britanak2007discrete}.
Typically images satisfy these requirements~\cite{britanak2007discrete}. 
The MSE
is mathematically
detailed in~\cite{Liang2001,britanak2007discrete}
and should be minimized to maintain the compatibility 
between 
the approximation and the \textit{exact} 
DCT outputs~\cite{britanak2007discrete}.

\subsubsection{Transform-related Measures}

The transform coding gain is
an important figure of merit 
to evaluate the coding efficiency of
a given transform
as a data compression tool.
Its mathematical description can be found 
in~\cite{Liang2001,britanak2007discrete}.

Another measure to evaluate the transform coding gain is the
transform efficiency~\cite{Cham1989,britanak2007discrete}.
The optimal KLT converts signals into 
completely uncorrelated coefficients that has 
transform efficiency equal to 100,
whereas the DCT-II achieves a transform efficiency of 93{.}9911
for Markovian data at 
correlation coefficient of 0{.}95~\cite{britanak2007discrete}.

\subsubsection{Image Quality Measures in JPEG-like Compression}

For quality analysis,
images were
submitted to
a
JPEG-like technique for image compression.
The resulting compressed images
are then assessed for image degradation
in comparison to the original input image.
Thus,
\mbox{2-D} versions of the discussed
methods are required.
An 8$\times$8 image block $\mathbf{A}$ has its
\mbox{2-D} transform mathematically expressed by~\cite{suzuki2010integer}:
\begin{align}
\label{equation-2d-transform}
\mathbf{T} \cdot \mathbf{A} \cdot \mathbf{T}^\top
,
\end{align}
where $\mathbf{T}$ is a considered transformation.
Input images
were
divided into $8 \times 8$ sub-blocks,
which were submitted to the \mbox{2-D} transforms. 
For each block,
this computation furnished 
64~coefficients in the approximate transform domain
for a particular transformation.
According to the standard zigzag sequence~\cite{pao1998},
only the 
$2 \leq r \leq 20$ initial coefficients in each block 
were retained 
and
employed for image reconstruction~\cite{cintra2011dct}. 
This range of $r$ corresponds to high compression.
All the remaining coefficients were set to zero.
The inverse procedure
was then applied to reconstruct the processed image.

Subsequently,
recovered images had their
PSNR~\cite{Salomon2007} and UQI~\cite{wang2002universal} evaluated. 
The PSNR is a standard quality metric
in the image processing literature~\cite{Quan2012},
and 
the UQI
is regarded as a more sophisticate tool
for quality assessment,
which takes into consideration structural similarities
between images under analysis~\cite{cintra2011dct,wang2002universal}.

This methodology was
employed in~\cite{haweel2001square}, 
supported in~\cite{bouguezel2008low, bouguezel2008multiplication, bouguezel2009fast, bouguezel2010novel,bouguezel2011parametric}, 
and extended in~\cite{cintra2011dct,cb14}. 
However,
in contrast to the 
JPEG-like experiments described 
in~\cite{bouguezel2008low, bouguezel2008multiplication, bouguezel2009fast, bouguezel2010novel,bouguezel2011parametric}, 
the extended experiments considered 
in~\cite{cintra2011dct,cb14}
adopted 
the average image quality measure from 
a collection of representative images
instead of 
resorting to measurements obtained from single particular images. 
This approach is less prone to 
variance effects and fortuitous input data, 
being more robust~\cite{kay1993}.
For the above procedure,
we considered 
a set of 45 8-bit greyscale $512 \times 512$
standard 
images obtained from a public image bank~\cite{uscsipi}.

\subsection{Performance Results}

\begin{figure}
\centering

\subfigure[Average PSNR]
{\includegraphics[width=0.6\linewidth]{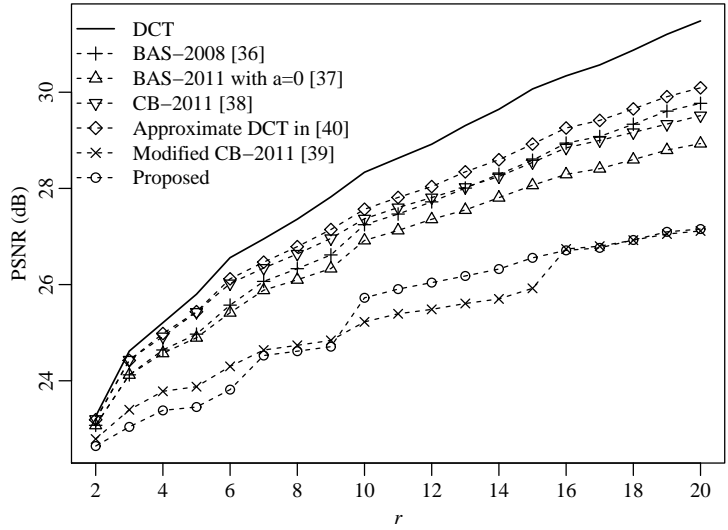}}
\subfigure[Average UQI absolute percentage error relative to the DCT]
{\includegraphics[width=0.6\linewidth]{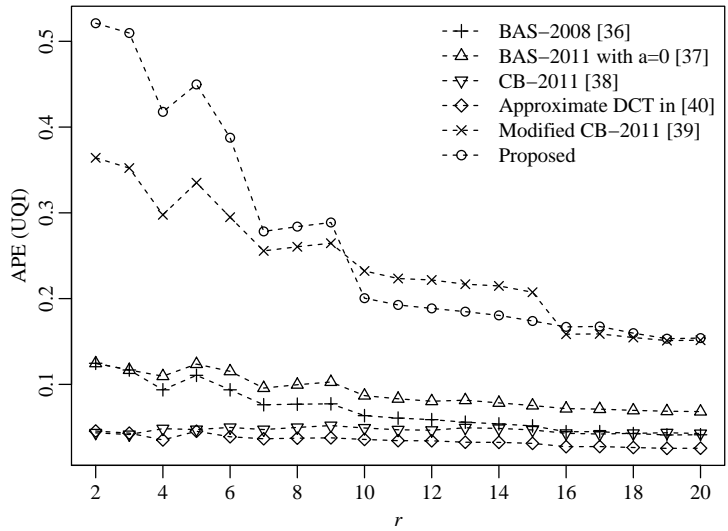}}

\caption{Image quality measures for several compression ratios.}
\label{figure-quality}
\end{figure}

Figure~\ref{figure-quality} presents
the resulting average PSNR and 
average UQI absolute percentage error~(APE) 
relative to the DCT, for $r=2,3,\ldots, 20$,
i.e., for high compression ratios~\cite{cintra2011dct}.
The proposed transform 
could outperform
the Modified \mbox{CB-2011} approximation  
for $10 \leq r \leq 15$, 
i.e.,
when 
84.38\%
to 
76.56\%
of the DCT coefficients are discarded.
Such high compression ratios 
are employed in several applications~\cite{bouguezel2008multiplication, 
Pan1999,Pan2001,Robinson2003,Mandyam1997}.

Table~\ref{table-performance} shows the 
performance measures for the considered transforms.
Average 
PSNR and UQI measures 
are presented for all considered images 
at a selected high compression ratio $r=10$.
The approximate transform 
proposed in~\cite{multibeam2012}
could outperform
remaining methods
in terms of proximity measures (total energy error and MSE)
when compared to the \emph{exact} DCT.
It also furnished
good image quality measure results 
(average $\text{PSNR}=27.567\,\mathrm{dB}$).
However,
at the same time, 
it is the most expensive approximation
measured by its computational cost
as shown in Table~\ref{table-complexity}.

On the other hand,
the transforms with lowest 
arithmetic complexities
are the 
Modified \mbox{CB-2011} approximation
and
new proposed transform,
both 
requiring only 14~additions.
The new transform
could 
outperform
the Modified \mbox{CB-2011} approximation
as an image compression tool
as indicated by the PSNR and UQI values.

\begin{figure}%
\centering
\subfigure[BAS-2008]
{\includegraphics[width=0.32\linewidth]{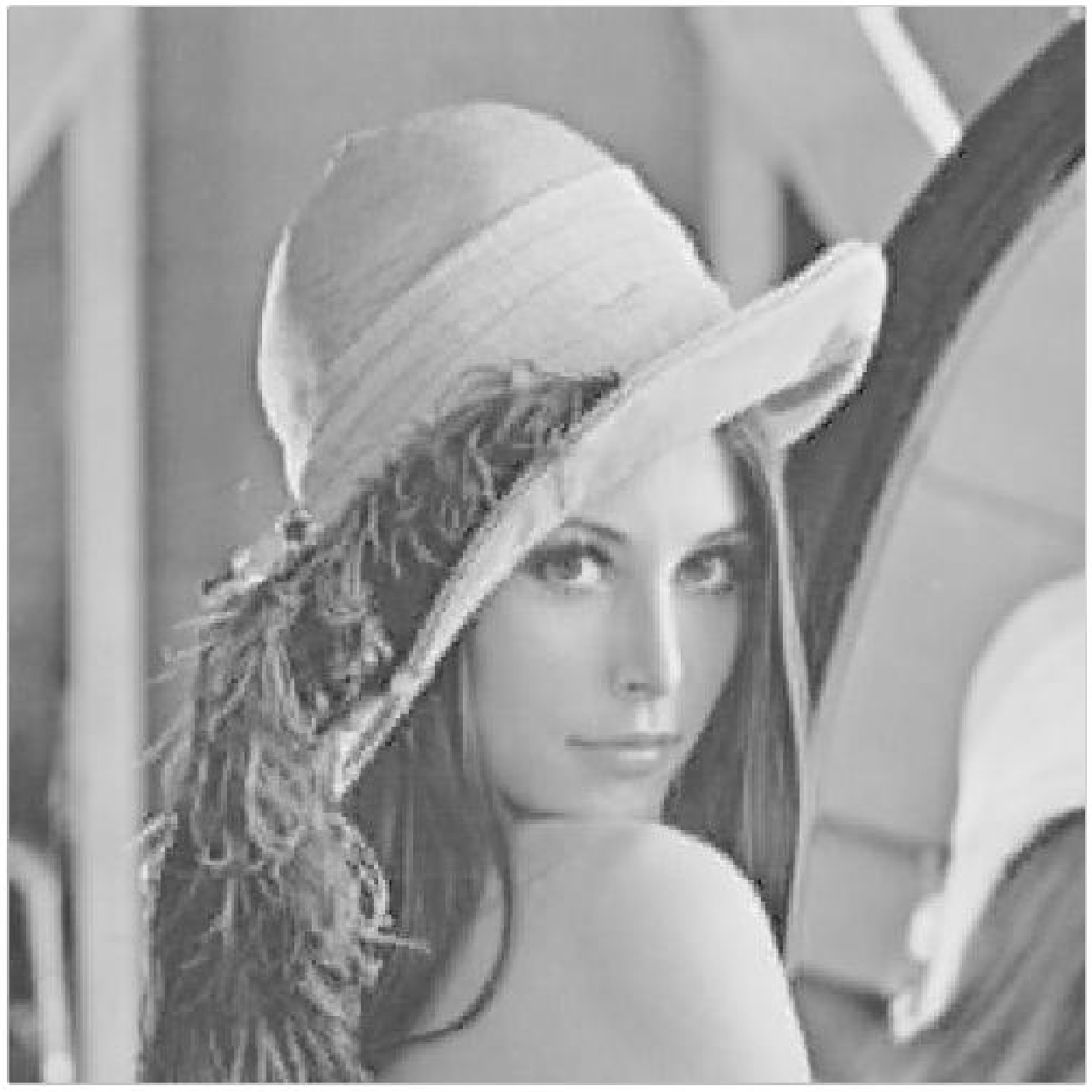}}
\subfigure[BAS-2011 ($a=0$)]
{\includegraphics[width=0.32\linewidth]{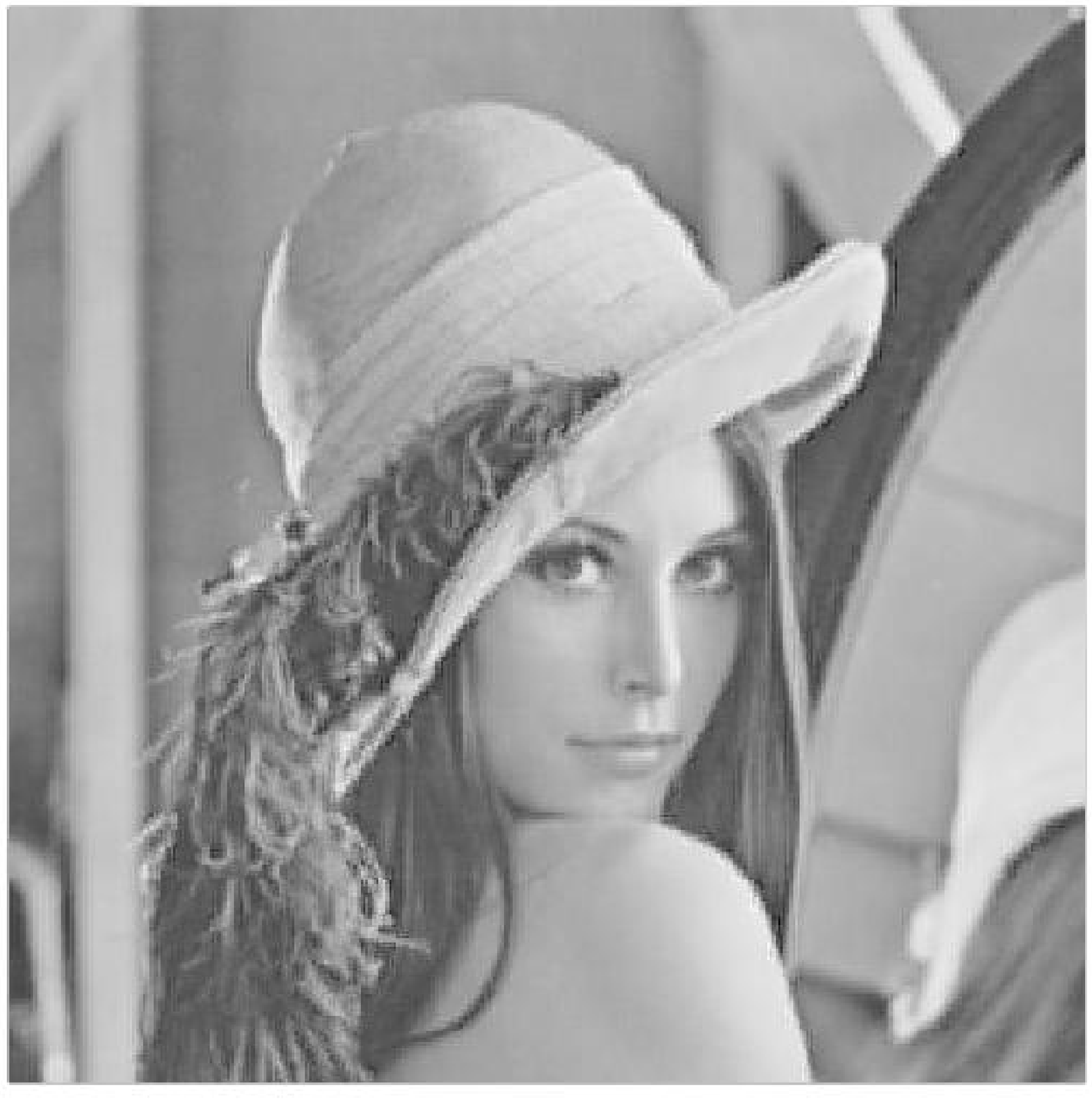}}
\subfigure[CB-2011]
{\includegraphics[width=0.32\linewidth]{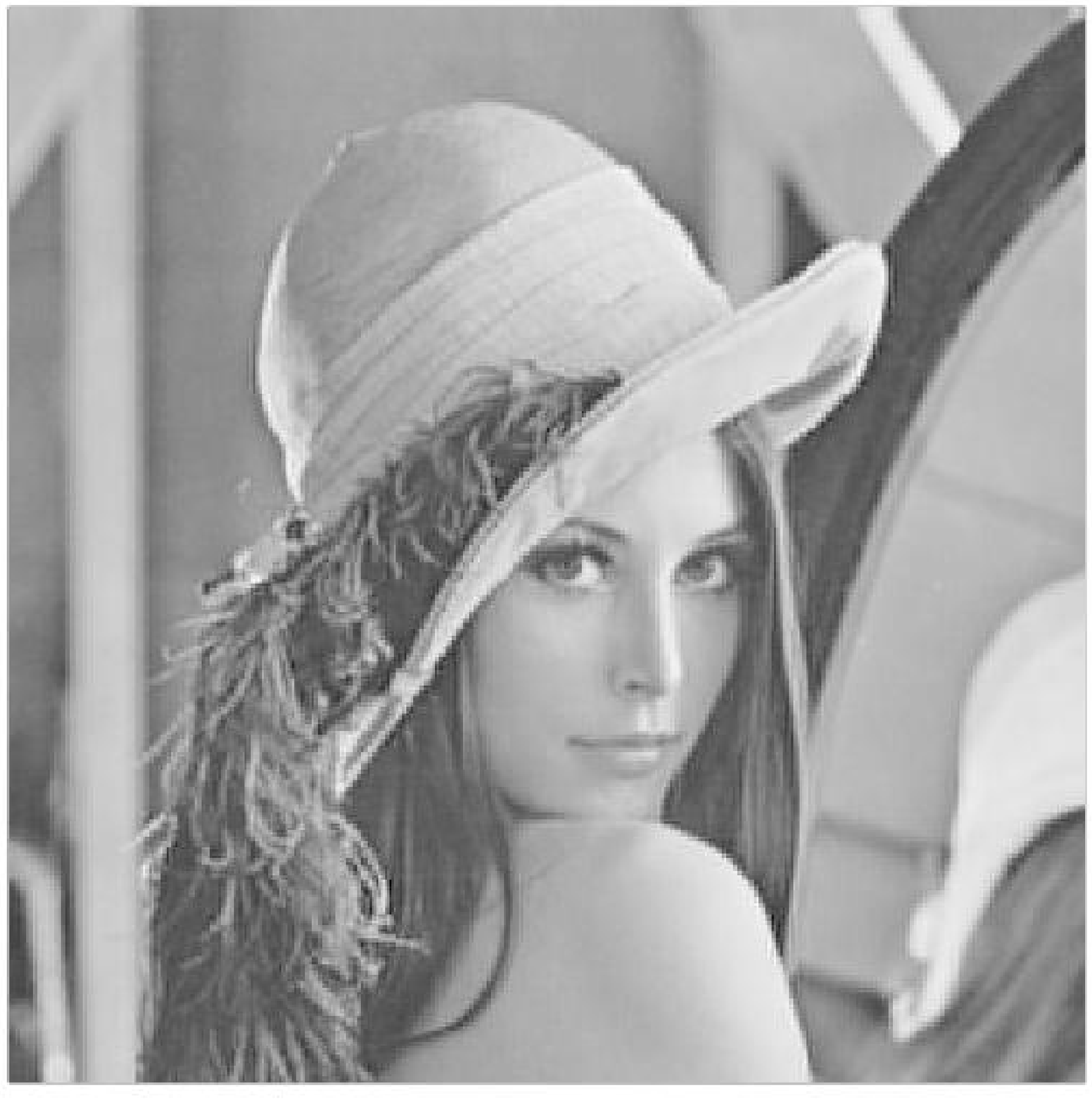}}
\subfigure[Modified \mbox{CB-2011}~\cite{cb14}]
{\includegraphics[width=0.32\linewidth]{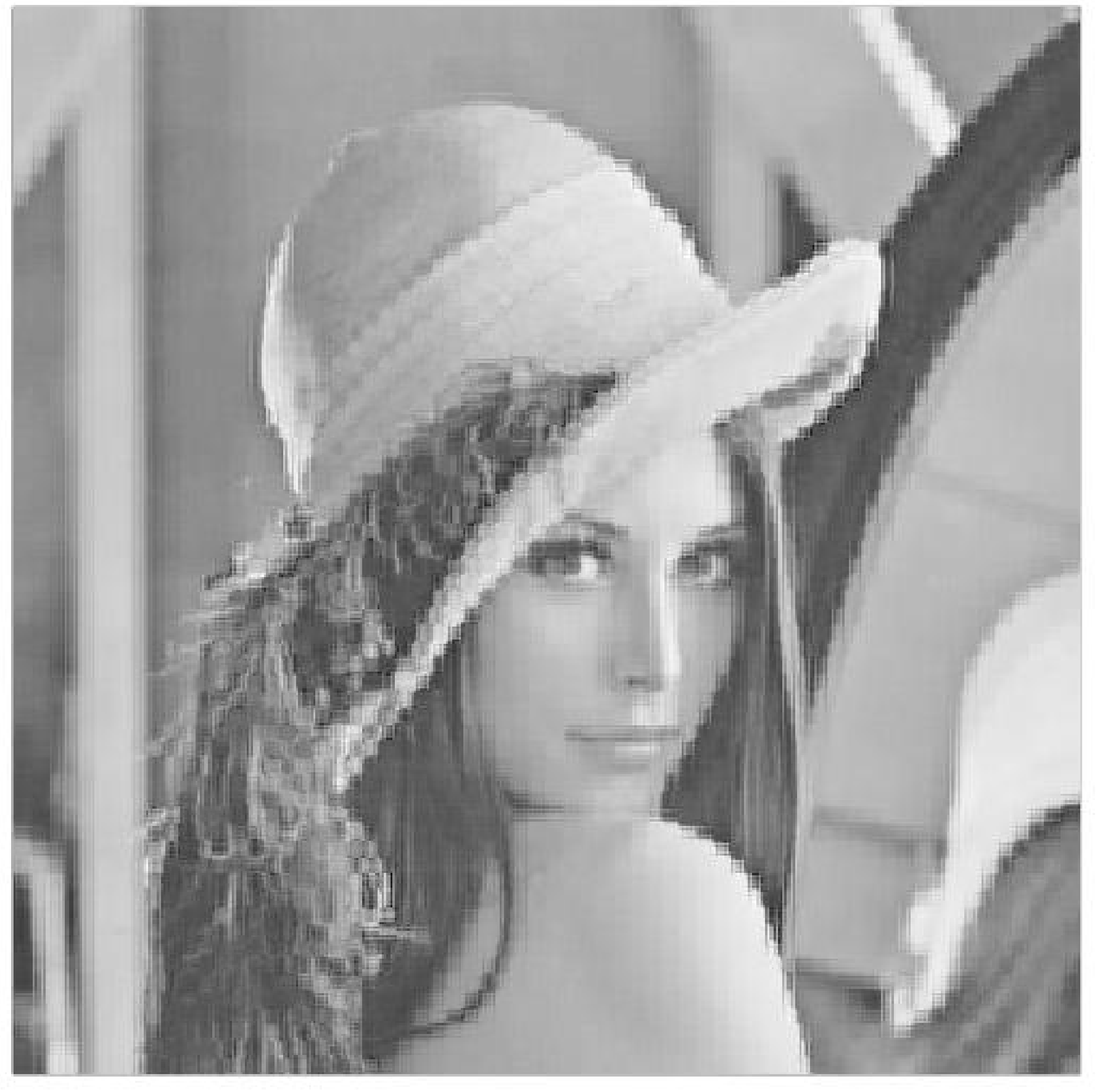}}
\subfigure[Approximate DCT in~\cite{multibeam2012}]
{\includegraphics[width=0.32\linewidth]{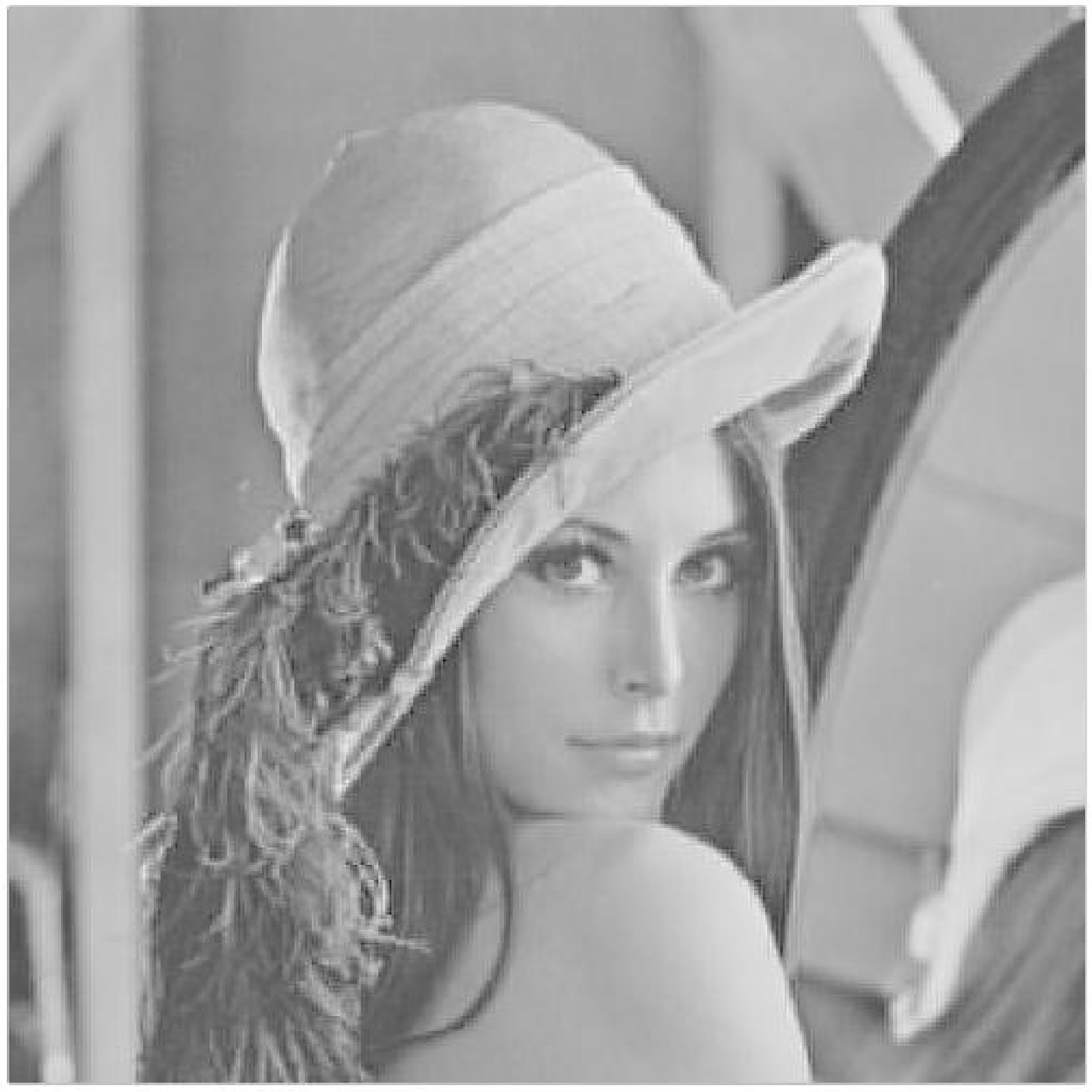}}
\subfigure[Proposed transform]
{\includegraphics[width=0.32\linewidth]{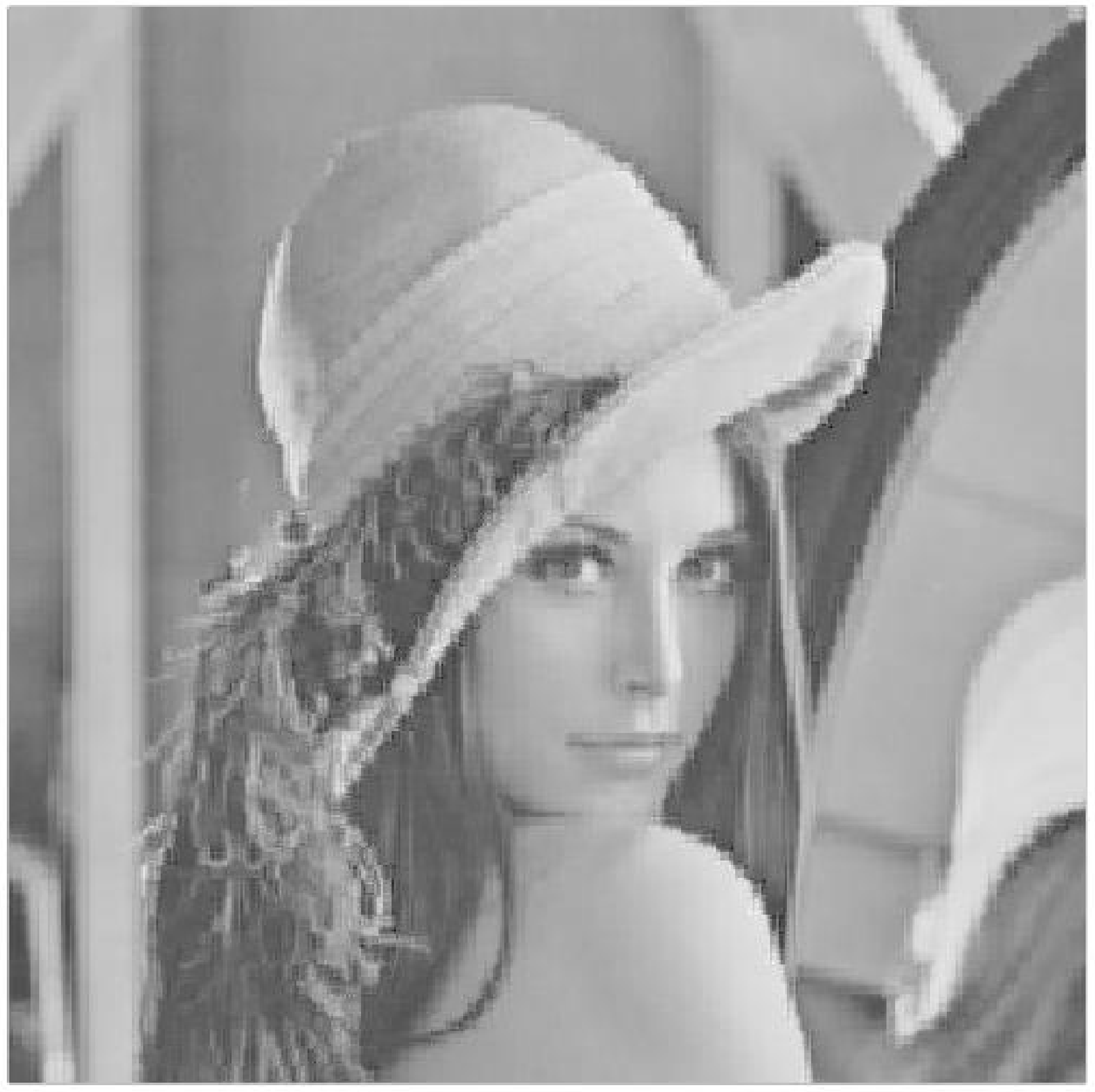}}

\caption{Compressed Lena image using several DCT approximations.
Compression ratio is $84.375\%$ ($r=10$).}
\label{lena}
\end{figure}

A qualitative comparison based on the resulting
compressed image Lena~\cite{uscsipi}
obtained from the above describe procedure
for $r=10$
is shown in Fig.~\ref{lena}.

\begin{table*}

\centering

\caption{Accuracy measures of discussed methods}

\label{table-performance}

\begin{tabular}{lcccccc} %
\toprule
Method
& 
$\epsilon$
& 
$\operatorname{MSE}$ ($\times \!10^{-2}$)
&
$C_g$  
& 
$\eta$ 
& 
PSNR
& 
UQI
\\
\midrule
\emph{Exact} DCT & 
0.000 & 0.000 & 8.826 & 93.991 & 28.336 & 0.733\\
BAS-2008~\cite{bouguezel2008low} & 
5.929 & 2.378 & 8.120 & 86.863 & 27.245 & 0.686\\
\mbox{BAS-2011}~\cite{bouguezel2011parametric} with $a=0$ & 
26.864 & 7.104 & 7.912 & 85.642 & 26.918 & 0.669\\
\mbox{BAS-2011}~\cite{bouguezel2011parametric} with $a=1$ & 
26.864 & 7.102 & 7.913 & 85.380 & 26.902 & 0.668\\
\mbox{BAS-2011}~\cite{bouguezel2011parametric} with $a=2$ & 
27.922 & 7.832 & 7.763 & 84.766 & 26.299 & 0.629\\
\mbox{CB-2011}~\cite{cintra2011dct} & 
1.794 & 0.980 & 8.184 & 87.432 & 27.369 & 0.697\\
Modified \mbox{CB-2011}~\cite{cb14} & 
8.659 & 5.939 & 7.333 & 80.897 & 25.224 & 0.563\\
Approximate DCT in~\cite{multibeam2012} & 
0.870 & 0.621 & 8.344 & 88.059 & 27.567 & 0.701\\
Proposed transform & 
11.313 & 7.899 & 7.333 & 80.897 & 25.726 & 0.586\\
\bottomrule
\end{tabular}

\end{table*}

Fig.~\ref{figure-quality} 
and 
Table~\ref{table-performance}
illustrate the usual trade-off
between
computational complexity and performance.
For instance,
although
BAS-2011 (for $a=0$) 
could yield a better PSNR figure
when
compared with the proposed algorithm,
it is computationally more demanding (about 14.3\% more operations)
and
its 
coding gain and transform efficiency 
are improved in only
7.9\% and 5\%, 
respectively.
In contrast,
the proposed algorithm requires only 14 additions,
which
can lead to smaller,
faster and
more energy efficient
circuitry designs.
In the next section,
we 
offer
a comprehensive hardware analysis
and comparison
of the discussed algorithms
with several
implementation specific
figures of merit.

We also notice that
although
the proximity of the exact DCT---as measured by the MSE---is 
a good characteristic, 
it is not the defining property 
of a good DCT approximation,
specially in image compression applications. 
A vivid example of this---seemingly counter-intuitive phenomenon---is
the BAS series of DCT approximation.
Such approximations
possess comparatively large values of proximity measures (e.g., MSE)
when compared with the exact DCT matrix.
Nevertheless,
they exhibit very good performance
 in image compression application. 
Results displayed in Table II illustrate this behavior.

\subsection{Implementation in Real Time Video Compression Software}

\begin{figure}
\centering
{\includegraphics[width=0.6\linewidth]{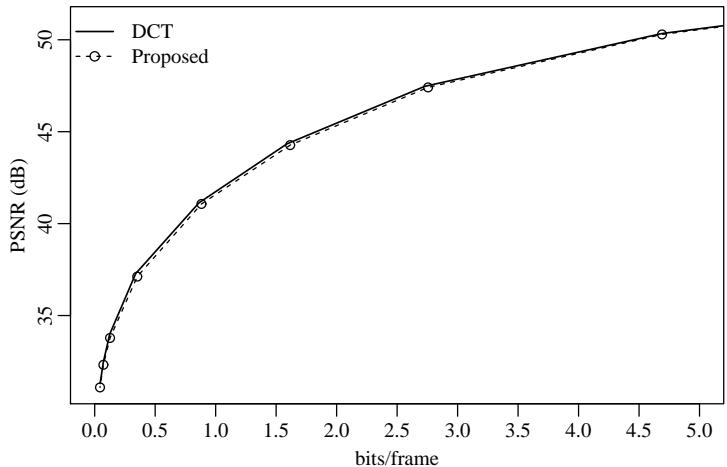}}
\caption{RD curves for `BasketballPass' test sequence.}\label{fig:RD}
\end{figure}

\begin{figure}
\centering
\subfigure[Chen DCT ($\text{QP}=0$)]
{\includegraphics[width=0.47\linewidth]{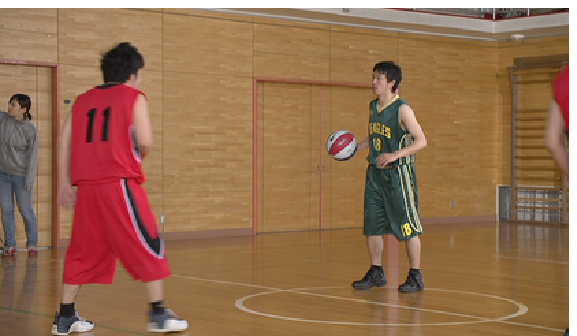}}
 \label{Chen QP0}
\subfigure[Proposed DCT ($\text{QP}=0$)]
{\includegraphics[width=0.47\linewidth]{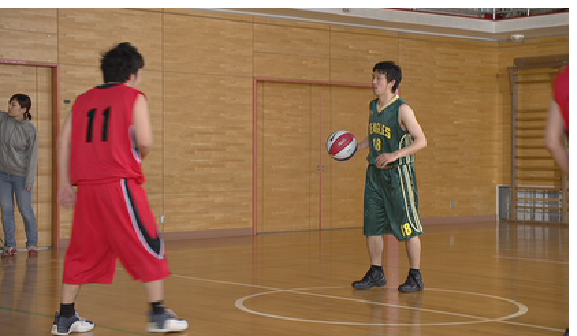}}
 \label{Proposed QP0}
\\*
\subfigure[Chen DCT ($\text{QP}=32$)]
{\includegraphics[width=0.47\linewidth]{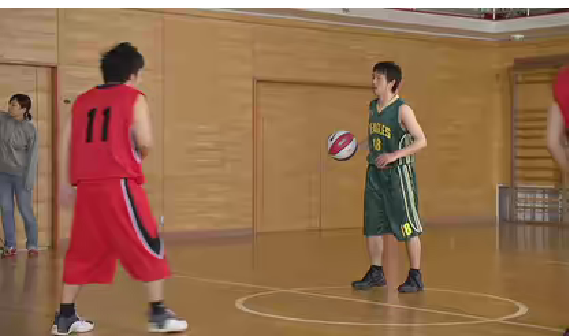}}
 \label{Chen QP32}
\subfigure[Proposed DCT ($\text{QP}=32$)]
{\includegraphics[width=0.47\linewidth]{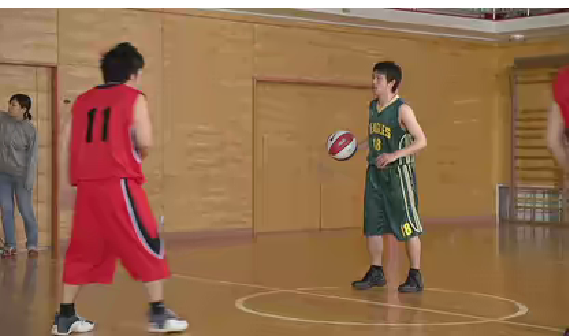}}
 \label{Proposed QP32}
\\*
 \subfigure[Chen DCT ($\text{QP}=50$)]
{\includegraphics[width=0.47\linewidth]{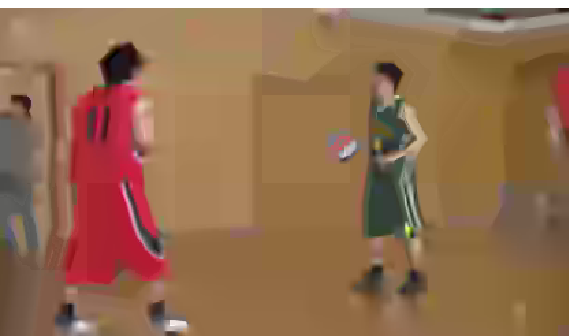}}
 \label{Chen QP50}
\subfigure[Proposed DCT ($\text{QP}=50$)]
{\includegraphics[width=0.47\linewidth]{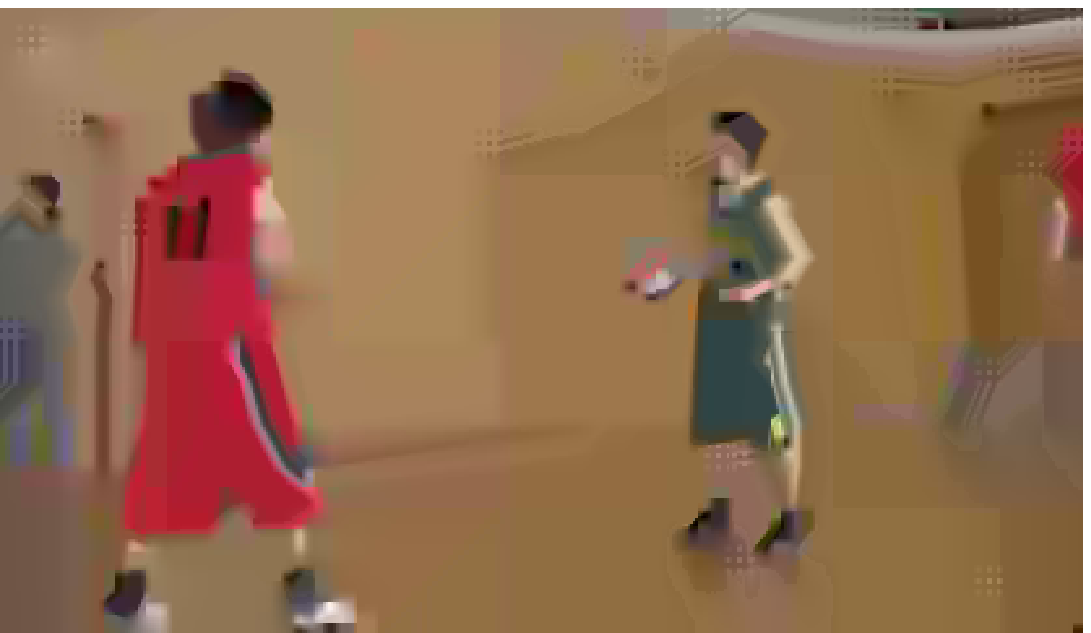}}
 \label{Proposed QP50}
 \caption{Selected frames from `BasketballPass' test video coded
by means of the Chen DCT and the proposed DCT approximation
for 
$\text{QP}= 0$ (a-b), 
$\text{QP} = 32$ (c-d),
and 
$\text{QP} = 50$ (e-f).}
\label{F:frames}
\end{figure}

The proposed approximate DCT transform was 
embedded into
an open source HEVC standard reference software~\cite{refsoft}
in order to assess its performance in real time video coding. 
The original integer 
transform prescribed in the selected HEVC reference software
is a scaled approximation of Chen DCT algorithm~\cite{chens},
which employs 26 additions.
For comparison,
the proposed approximate DCT requires only 14 additions.
Both algorithms
were evaluated for their effect on 
the overall performance of the encoding process
by obtaining rate-distortion~(RD) curves
for standard video sequences.
The curves were obtained by varying 
the quantization point (QP) 
from 0 to 50 and obtaining the PSNR of 
the proposed
approximate transform with reference to 
the Chen DCT implementation,
which is already implemented in the reference software, 
along with the bits/frame of the encoded video.
The PSNR computation was performed by taking the average
PSNR obtained from the three channels YCbCr of the color image,
as suggested in~\cite[p.~55]{semary2012image}.
Fig.~\ref{fig:RD} depicts the obtained RD curves 
for the `BasketballPass' test sequence. 
Fig.~\ref{F:frames} shows particular
416$\times$240 frames
for
$\text{QP} \in \{0, 32, 50\}$
when
the proposed approximate DCT and
the Chen DCT
are considered.

The RD curves reveals that 
the difference in the rate points of Chen DCT and proposed approximation 
is negligible. 
In fact, 
the mean absolute difference was 0.1234~dB,
which is very low.
Moreover,
the frames show 
that 
both encoded video streams using the above two DCT transforms are almost identical. 
For each QP value,
the PSNR values between the resulting frames
were
82.51~dB,
42.26~dB,
and
36.38~dB,
respectively.
These very high PSNR values confirm the adequacy of the proposed scheme.

\section{Digital Architectures and Realizations}
\label{section-hardware}

In this section
we propose architectures for the detailed \mbox{1-D}
and \mbox{2-D} approximate 8-point DCT.
We aim at physically implementing~\eqref{equation-2d-transform}
for various transformation matrices.
Introduced architectures were submitted to
(i)~Xilinx FPGA implementations
and 
(ii)~CMOS 45~nm application specific integrated circuit (ASIC) implementation up to the synthesis level.

This section explores the hardware utilization of 
the discussed algorithms
while providing a comparison with 
the proposed novel DCT approximation algorithm and its fast algorithm realization.
Our objective here is to offer digital realizations together with
measured or simulated metrics of hardware resources so that better decisions on the choice of a particular fast algorithm and its implementation can be reached.

\subsection{Proposed Architectures}

\begin{figure}
\centering
\scalebox{1.15}{\input{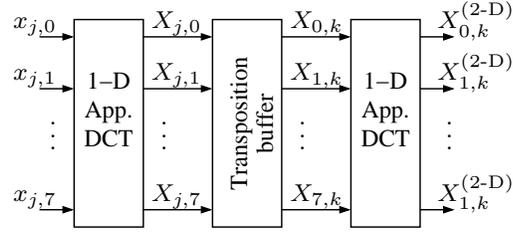}}
\caption{%
Two-dimensional approximate transform by means of \mbox{1-D} approximate transform.
Signal $x_{k,0},x_{k,1},\ldots$ corresponds to the rows of the input image;
$X_{k,0},X_{k,1},\ldots$ indicates the transformed rows;
$X_{0,j},X_{1,j},\ldots$ indicates the columns of
the transposed row-wise transformed image;
and
$X_{0,j}^\text{(2-D)},X_{1,j}^\text{(2-D)},\ldots$ indicates the columns of
the final \mbox{2-D} transformed image.
If $i=0,1,2,3,\ldots$, 
then indices $j$ and $k$ satisfy 
$j= i \pmod{8}$ and $k=[(\downarrow8)i]/8 \pmod{8}$.}
\label{2D_arch}
\end{figure}

We propose digital computer architectures that 
are custom designed for the real-time
implementation of the fast algorithms described in 
Section~\ref{section-review}. 
The proposed architectures
employs
two parallel realizations of DCT approximation blocks,
as shown in Fig.~\ref{2D_arch}.

The \mbox{1-D} approximate DCT blocks
(Fig.~\ref{2D_arch})
implement a particular fast algorithm chosen from 
the collection described earlier in the paper.
The first instantiation of the DCT block
furnishes a row-wise transform computation of the input image, 
while the second implementation furnishes a column-wise transformation
of the intermediate result.
The row- and column-wise transforms 
can be any of the DCT approximations detailed in the paper. 
In  other words, 
there is no restriction 
for both row- and column-wise transforms to be the same. 
However, for simplicity, 
we adopted identical transforms for both steps.

\begin{figure*}
\centering
{\input{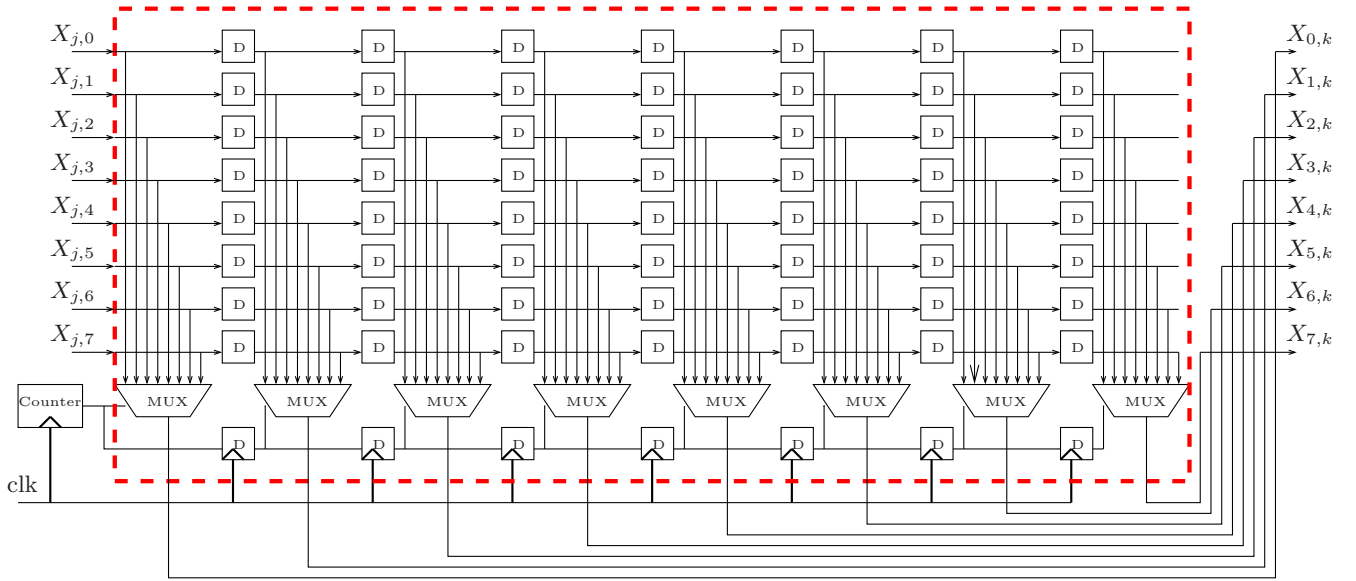}}
\caption{Details of the transposition buffer block.}
\label{da1}
\end{figure*}

Between the approximate DCT blocks 
a real-time row-parallel transposition buffer circuit
is required.
Such block ensures data ordering for converting the row-transformed data 
from the first DCT approximation circuit to 
a transposed format as required by the column transform circuit.
The transposition buffer block is detailed in~Fig.~\ref{da1}.

\begin{figure*}[htp]
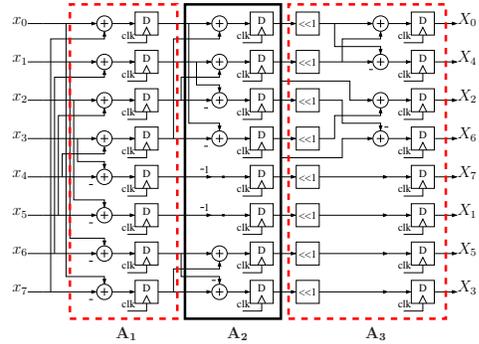
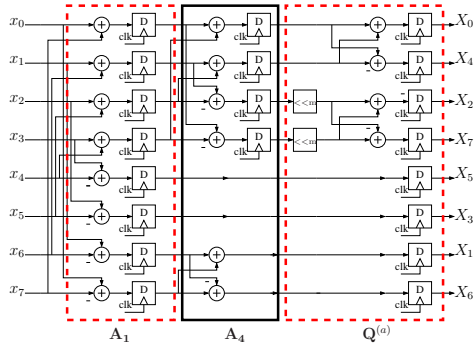
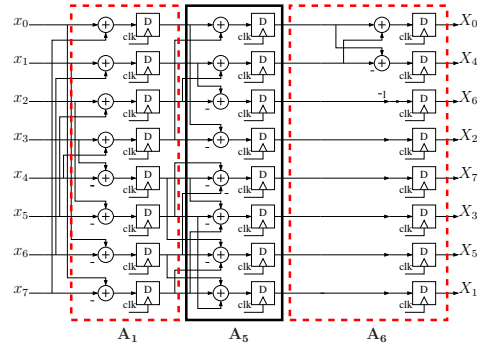
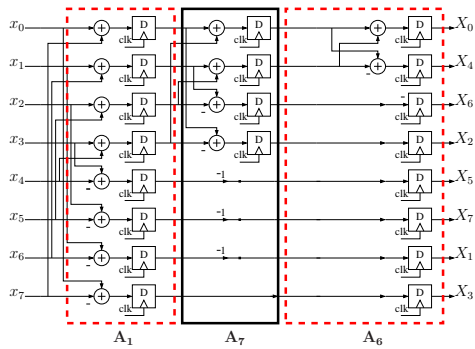
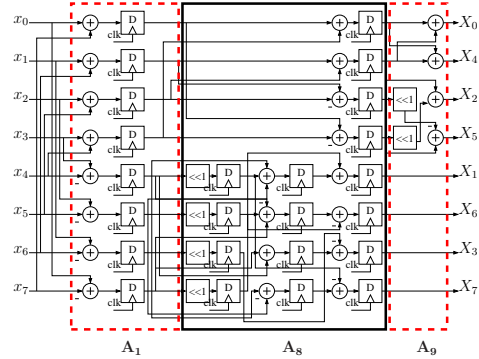

\centering
\subfigure[Proposed approximate transform ($\mathbf{T}^\ast$).]
{{\scalebox{0.51}{\input{prop2-circuit.pstex_t}}} \label{figure-pt3}}
\hspace{0.85cm}
\subfigure[BAS-2008 approximate DCT ($\mathbf{T}_1$).]
{{\scalebox{0.51}{\input{bas-circuit.pstex_t} } } \label{figure-bas}}\\
\subfigure[BAS-2011 approximate DCT ($\mathbf{T}^{(a)}$) where $m\in\{-\infty,0,1\}$.]
{{\scalebox{0.51}{\input{parametric-circuit.pstex_t}}} \label{figure-pbas} }
\hspace{0.85cm}
\subfigure[CB-2011 approximate DCT ($\mathbf{T}_2$).]
{{\scalebox{0.51}{\input{cb-circuit.pstex_t}}} \label{figure-cb}}\\
\subfigure[Modified CB-2011 approximate DCT ($\mathbf{T}_3$).]
{{\scalebox{0.51}{\input{prop3-circuit.pstex_t}}} \label{figure-pt2}}
\hspace{0.85cm}
\subfigure[Approximate DCT in~\cite{multibeam2012} ($\mathbf{T}_4$).]
{{\scalebox{0.51}{\input{prop1-circuit.pstex_t}}} \label{figure-pt1}}
\caption{Digital architecture for considered DCT approximations.}
\label{architecture}
\end{figure*}

The digital architectures of the discussed 
approximate DCT algorithms 
were given
hardware signal flow diagrams
as listed below:

\begin{enumerate}

\item 
Proposed novel algorithm and architecture 
shown in Fig.~\ref{figure-pt3};

\item 
BAS-2008  architecture 
shown in Fig.~\ref{figure-bas};

\item 
BAS-2011 architecture 
shown in Fig.~\ref{figure-pbas};

\item 
CB-2011  architecture 
shown in Fig.~\ref{figure-cb};

\item 
Modified CB-2011 architecture 
shown in Fig.~\ref{figure-pt2};

\item 
Architecture for the algorithm in~\cite{multibeam2012} 
shown in Fig.~\ref{figure-pt1}.

\end{enumerate}
The circuitry sections associated to the
constituent matrices of the discussed factorizations
are emphasized in the figures
in bold or dashed boxes.

\subsection{Xilinx FPGA Implementations} 

Discussed methods
were 
physically realized on a FPGA based
rapid prototyping system for 
various register sizes and 
tested using on-chip hardware-in-the-loop co-simulation.
The architectures were designed for 
digital realization within the MATLAB environment
using the Xilinx System Generator (XSG) with 
synthesis options set to generic VHDL generation. 
This was necessary because the auto-generated register transfer language (RTL) 
hardware descriptions are targeted on both FPGAs as well as 
custom silicon using standard cell ASIC technology.

The proposed architectures were physically realized on 
Xilinx Virtex-6 XC6VSX475T-2ff1156 device.
The architectures were realized with 
fine-grain pipelining for increased throughput.
Clocked registers were inserted at appropriate points within 
each fast algorithm in order 
to reduce the critical path delay as much as possible at 
a small cost to total area. 
It is expected that the additional logic overheard 
due to fine grain pipelining is marginal. 
Realizations were verified on FPGA chip using a Xilinx~ML605 board at 
a clock frequency of 100~MHz.
Measured results from the FPGA realization 
were achieved using stepped hardware-in-the-loop verification.  

Several input precision levels were considered 
in order to investigate the performance 
in terms of digital logic resource consumptions 
at varied degrees of numerical accuracy and dynamic range.
Adopting
system word length $L \in \{4,8,12,16 \}$,
we applied 10,000 random 8-point input test vectors using 
hardware co-simulation.
The test vectors were generated from within the MATLAB environment
and 
routed to the physical FPGA device using JTAG~\cite{fpga1}
based hardware co-simulation.
JTAG is a digital communication standard for 
programming and debugging reconfigurable devices such as Xilinx FPGAs.

Then the measured data from the FPGA was routed back to MATLAB memory space. 
Each FPGA implementation was evaluated 
for hardware complexity and real-time performance using metrics 
such as 
configurable logic blocks (CLB) and flip-flop (FF) count, 
critical path delay ($T_\text{cpd}$) in ns,
and maximum operating frequency ($F_\text{max}$) in MHz.
The number of available CLBs and FFs were 297{,}600 and 595{,}200,
respectively.

Results are reported in Table~\ref{fpga}.
Quantities were obtained from the Xilinx FPGA synthesis
and place-route tools by 
accessing the \texttt{xflow.results} report file for each run of the design flow.
In addition, the static ($Q_p$) and dynamic power ($D_p$) consumptions were estimated using the Xilinx XPower Analyzer.

From Table~\ref{fpga} it is evident that the proposed transform and the modified CB-2011 approximation are faster than remaining
approximations.
Moreover,
these two particular designs
achieve the lowest consumption of hardware resources
when compared with remaining designs.

\begin{table}
\centering
\tablesize
\caption{Hardware Resource Consumption using Xilinx Virtex-6 XC6VSX475T-2ff1156 device}
\label{fpga}
\begin{tabular}{ccccccc} %

\toprule

$L$ & 
\parbox{1cm}{\centering CLB} &
\parbox{1cm}{\centering FF } &
$Q_p$ ($\mathrm{W}$) & 
$D_p$ ($\mathrm{W}$) &
$T_\text{cpd}$  & 
$F_{\text{max}}$
\\ [0.5ex] %

\midrule
\multicolumn{7}{c}{\centering BAS-2008 Algorithm} \\
\midrule

4 & 395&	784&	5.154&	0.918&	2.350&	401.7\\
8 & 613	&1123	&5.168&	1.105	&2.573&	367.1 \\
12 & 821&	1523&	5.184&	1.301&	2.930&	337.8 \\
16&1029	&1915&	5.187&	1.344	&3.254&	284.0\\

\midrule
\multicolumn{7}{c}{BAS-2011 for $a=0$} \\
\midrule

4 & 335&	877&	5.142&	0.767&	2.340&	386.4\\
8 & 535&1276	&5.161&	1.015	&2.600&	356.2 \\
12 & 728&	1732&	5.180&	1.260&	2.822&	337.4 \\
16&919	&2187&	5.198&	1.486	&2.981&	325.2\\

\midrule
\multicolumn{7}{c}{BAS-2011 for $a=1$} \\
\midrule

4 & 387&	1019&	5.146&	0.811&	2.413&	396.7\\
8 & 605	&1453	&5.165&	1.065	&2.513&	361.4 \\
12 & 813&	1949&	5.179&	1.247&	2.962&	329.4 \\
16&1021	&2445&	5.198&	1.483	&2.987&	316.9\\

\midrule
\multicolumn{7}{c}{BAS-2011 for $a=2$} \\
\midrule

4 & 385&	1019&	5.146&	0.818&	2.371&	402.9\\
8 & 603	&1453	&5.163&	1.042	&2.584&	364.7 \\
12 & 812&	1950&	5.190&	1.378&	2.618&	353.1 \\
16&1019	&2445&	5.201&	1.527	&3.006&	326.5\\

\midrule
\multicolumn{7}{c}{CB-2011 Algorithm} \\
\midrule

4 & 452&	883&	5.141&	0.750&	2.518&	363.4 \\
8 & 702	&1257	&5.151&	0.876	&3.065&	303.1 \\
12	&950	&1709	&5.162&	1.029&	3.466&	270.6\\
16	&1198	&2162&	5.187&	1.341&	3.610&	256.0\\

\midrule
\multicolumn{7}{c}{Approximate DCT in~\cite{multibeam2012}} \\
\midrule
4 & 513	&1040&	5.158&	0.972&	2.545&	387.8 \\
8 & 779	&1471&	5.173	&1.170&	2.769&	351.0\\
12	&1036	&1968&	5.181	&1.262&	2.945&	314.9\\
16&	1291&	2463&	5.200&	1.514&	3.205&	298.0 \\

\midrule
\multicolumn{7}{c}{Modified CB-2011 approximation} \\
\midrule

4 &297&	652&	5.153&	0.903&	2.384&	399.7\\
8 & 481&	961	&5.177&	1.214&	2.523&	391.2\\
12	&657&	1329&	5.191&	1.390&	2.693&	354.0\\
16	&834	&1698&	5.219&	1.752&	2.829&	345.5\\

\midrule
\multicolumn{7}{c}{Proposed Transform} \\
\midrule

4 &303&	651&	5.146&	0.818&	2.344&	404.0\\
8 & 487	&963	&5.167&	1.092&	2.470&	385.1\\
12	&663&	1329	&5.185&	1.322	&2.524&	353.7\\
16	&839&	1697	&5.203&	1.551&	2.818&	341.8\\

\bottomrule
\end{tabular}
\end{table}

\subsection{CMOS 45~nm ASIC Implementation}

The digital architectures were first designed using 
Xilinx System Generator tools within the Matlab/Simulink environment. 
Thereafter, 
the corresponding circuits were simulated using bit-true cycle-accurate models 
within the Matlab/Simulink software framework. 
The architectures were then converted to 
corresponding digital hardware description language designs using
the auto-generate feature of the System Generator tool.
The resulting hardware description language code led to 
physical implementation of the architectures using 
Xilinx FPGA technology, 
which in turn led to extensive hardware co-simulation on FPGA chip.
Hardware co-simulation was used for verification of 
the hardware description language designs 
which were contained in register transfer language (RTL) libraries.
Thus, 
the above mentioned verified RTL code 
for each of the \mbox{2-D} architectures was ported to
the Cadence RTL Compiler environment 
for mapping to application specific CMOS technology. 
To guarantee that the auto-generated RTL 
could seamlessly compile in the CMOS design environment, 
we ensured that RTL code followed a behavioral description 
which did not contain any FPGA specific (vendor specific) instructions. 
By adopting standard IEEE~1164 libraries and behavioral RTL, 
the resulting code was compatible with Cadence Encounter for 
CMOS standard cell synthesis. 

For this purpose, 
we used FreePDK, 
a free open-source ASIC standard cell library at the 45~nm node~\cite{FPDK}.
The supply voltage of the CMOS realization was fixed at 
$V_\text{DD} = 1.1~\mathrm{V}$ 
during estimation of power consumption and logic delay. 
The adopted figures of merit for the ASIC synthesis
were:
area ($A$) in~$\mathrm{mm^2}$, 
critical path delay ($T$) in~ns, 
area-time complexity ($AT$) in $\mathrm{mm^2 \cdot ns}$, 
dynamic power consumption in watts,
and area-time-squared complexity ($AT^2$) in $\mathrm{mm^2 \cdot ns^2}$.
Results are displayed in Table~\ref{cad} and~\ref{cad-power}.

The $AT$ complexity is an adequate metric when 
the chip area is more relevant than speed or computational throughput. 
On the other hand, 
$AT^2$ is employed when real-time speed is the most important driving force
for the optimizations in the logic designs. 
In all cases,
clear improvements in maximum real-time clock frequency is predicted over the same RTL targeted at FPGA technology. 

\begin{table}
\centering
\tablesize
\caption{Hardware resource consumption for CMOS 45nm ASIC implementation}
\label{cad}
\begin{tabular}{ccccccc}

\toprule

\!\!\!$L$ \!\!\!\!\!& 
\!\!\!\parbox{0.5cm}{ASIC Gates} \!\!\!\!\!& 
\!\!\!Area \!\!\!\!\!&
\!\!\! $T_\text{cpd}$ \!\!\!\!\!& 
\!\!\!$AT$ \!\!\!\!\!& 
\!\!\!$AT^2$ \!\!\!\!\!&
\!\!\!$F_{\text{max}}$
\\ [0.5ex]

\midrule
\multicolumn{7}{c}{BAS-2008 Algorithm} \\
\midrule

4 & 27792& 0.123&	1.140& 0.140 &0.160 &	877.2 \\
8 & 44654 &0.192	&1.204 &0.231 &0.278	&830.6  \\
12 & 61388&0.262&	1.216&0.319 &0.388 &	822.4  \\
16&78281	& 0.332&1.236& 0.411&0.508 &	809.1 	\\

\midrule
\multicolumn{7}{c}{BAS-2011 for $a=0$} \\
\midrule
4 & 26299&0.114 &	1.135&	0.129& 0.147&881.1 \\
8 & 42313 &	0.182&1.147	&0.209 &0.239 &871.8  \\
12 & 58342 & 0.250&	1.225& 0.306& 0.375& 816.3 \\
16&74062 &	0.317&1.310&0.415 &0.544 &	763.4 	\\

\midrule
\multicolumn{7}{c}{BAS-2011 for $a=1$} \\
\midrule
4 & 25940&0.108 &	1.106&	0.120& 0.133&904.2 \\
8 & 40330 &	0.166&1.125	&0.187 &0.210 &888.9  \\
12 & 53728 & 0.225&	1.170& 0.263& 0.308& 	854.7  \\
16&67860 &	0.283&1.200&0.339 &0.407 &	833.3 	\\

\midrule
\multicolumn{7}{c}{BAS-2011 for $a=2$} \\
\midrule
4 & 25554& 0.109	&1.117&0.122	& 0.136&895.3 \\
8 & 39321 &0.167	&1.132	& 0.189&0.214 &883.4  \\
12 & 53950& 0.226&	1.175&	0.265&0.312 & 851.1  \\
16&67979 &	0.284&1.201&0.341 &0.409 & 	832.6  	\\

\midrule
\multicolumn{7}{c}{CB-2011 Algorithm} \\
\midrule

4 & 30319 &	0.132 & 1.167& 0.154	&0.180 & 856.9 \\
8 & 48556	&0.209  &1.192	&0.249 &0.296 &838.9  \\
12 & 66956&	0.285 & 1.221& 0.348&0.425 & 	819.0  \\
16& 85873 & 0.363	&1.240&	0.450 &0.558 &806.5 	\\

\midrule
\multicolumn{7}{c}{Approximate DCT in~\cite{multibeam2012}} \\
\midrule
4 & 35141 &0.151 &	1.141&0.173	&0.197 & 876.4 \\
8 & 53624&0.230	&1.211	& 0.278&0.337 &825.8  \\
12 & 73224 & 0.310&	1.234&0.383 &0.473 & 	810.4  \\
16& 92697 &	0.391&1.242& 0.486&0.603 & 	805.2 	\\

\midrule
\multicolumn{7}{c}{Modified CB-2011 Approximation} \\
\midrule
4 & 24777 &	0.107 & 1.105&0.119	&0.131 & 905.0 \\
8 & 40746	&0.175 &1.128&0.197 &0.222 	&886.5  \\
12 & 56644&	0.242& 1.164& 0.282&0.328 & 	859.1  \\
16& 73702 &0.314	&1.177 &0.369 &0.434  &849.6 \\

\midrule
\multicolumn{7}{c}{Proposed Transform} \\
\midrule
4 & 24817 &	0.107 & 1.110&0.119	&0.132 &900.9 \\
8 & 40705& 0.175	&1.129	&0.197 &0.223 &885.7  \\
12 & 56703 &	0.242& 1.165&0.282	&0.329 &858.4  \\
16& 73906& 	0.314&1.174 &0.368 &0.432&851.8	\\

\bottomrule
\end{tabular}
\end{table}

\begin{table}
\centering
\tablesize
\caption{Power consumption for CMOS 45nm ASIC implementation}
\label{cad-power}
\begin{tabular}{ccc}

\toprule

\!\!\!$L$ \!\!\!\!\!& 
\!\!\!$Q_p$($\mathrm{mW}$) \!\!\!\!\!&
$D_p$($\mathrm{W}$) 
\\ [0.5ex]

\midrule
\multicolumn{3}{c}{BAS-2008 Algorithm} \\
\midrule

4 & 	1.00&	0.18\\
8 &     1.56	&0.33 \\
12 &	2.13&	0.43 \\
16 & 2.70&	0.55	\\

\midrule
\multicolumn{3}{c}{BAS-2011 for $a=0$} \\
\midrule
4 & 0.94&	0.21\\
8 & 1.48	&0.33 \\
12 &	2.04&	0.42\\
16 & 2.59 &	0.50	\\

\midrule
\multicolumn{3}{c}{BAS-2011 for $a=1$} \\
\midrule
4 & 	0.88&	0.24\\
8 & 	1.34	&0.36 \\
12 & 	1.81&	0.40 \\
16 &    2.28&	0.48	\\

\midrule
\multicolumn{3}{c}{BAS-2011 for $a=2$} \\
\midrule
4 & 0.89&	0.20\\
8 & 1.35	&0.30 \\
12 &	1.82&	0.39 \\
16 & 2.29 &	0.48	\\

\bottomrule
\end{tabular}
\begin{tabular}{ccc}

\toprule

\!\!\!$L$ \!\!\!\!\!& 
\!\!\!$Q_p$($\mathrm{mW}$) \!\!\!\!\!&
$D_p$($\mathrm{W}$) 
\\ [0.5ex]

\midrule
\multicolumn{3}{c}{CB-2011 Algorithm} \\
\midrule

4 & 	1.08&	0.24\\
8 & 	1.706	&0.36 \\
12 & 	2.32&	0.48 \\
16 & 2.94 &	0.60	\\

\midrule
\multicolumn{3}{c}{Approximate DCT in~\cite{multibeam2012}} \\
\midrule
4 & 1.23&	0.28\\
8 & 1.87	&0.39 \\
12 & 2.52&	0.52 \\
16 &  3.17 &	0.64	\\

\midrule
\multicolumn{3}{c}{Modified CB-2011} \\
\midrule
4 & 	0.88&	0.20\\
8 & 	1.42	&0.32 \\
12 & 	1.98&	0.43 \\
16 & 2.55  &0.55		\\

\midrule
\multicolumn{3}{c}{Proposed Transform} \\
\midrule
4 &	0.88 &	0.20\\
8 & 	1.42	&0.32 \\
12 &    1.98&	0.43 \\
16 & 	 2.55 &0.55		\\

\bottomrule
\end{tabular}
\end{table}

\section{Conclusion}
\label{section-conclusion}

In this paper,
we proposed
(i)~a novel low-power 8-point DCT approximation that require only 14 addition operations to computations
and
(ii)~hardware implementation for the proposed transform and
several other prominent approximate DCT methods,
including the designs by Bouguezel-Ahmad-Swamy.
We obtained that all considered approximate transforms perform
very close to the ideal DCT. 
However, 
the modified \mbox{CB-2011} approximation and the proposed transform 
possess lower computational complexity
and
are faster than all other approximations under consideration. 
In terms of image compression, 
the proposed transform could outperform
the modified \mbox{CB-2011} algorithm.
Hence the new proposed transform is 
the best approximation for the DCT in terms of 
computational complexity and speed 
among the approximate transform examined.

Introduced implementations address both \mbox{1-D} and \mbox{2-D}
approximate DCT.
All the approximations were digitally implemented using both
Xilinx FPGA tools and CMOS 45~nm ASIC technology. 
The speeds of operation were much greater using the CMOS technology 
for the same function word size. 
Therefore,
the proposed architectures are suitable for image and video processing,
being candidates for improvements in several standards
including the HEVC.

Future work includes 
replacing the FreePDK standard cells with 
highly optimized proprietary digital libraries
from TSMC PDK~\cite{FPDK}
and
continuing the CMOS realization all the way up to chip fabrication 
and post-fab test on a measurement system. 
Additionally, 
we intend to develop the approximate versions for 
the 4-, 16-, and 32-point DCT
as well as to
the 4-point discrete sine transform,
which are
discrete transforms
required by HEVC.

\section{Acknowledgments}

This work was supported by
the University of Akron, Ohio, USA;
the 
\emph{Conselho Nacional de Desenvolvimento Cient\'ifico e Tecnol\'ogico} (CPNq) and FACEPE, 
Brazil;
and 
the Natural Science and Engineering Research Council (NSERC), 
Canada.

\bibliographystyle{IEEEtran}
\bibliography{references}

\end{document}